\documentclass[numreferences]{kluwer}
\usepackage[dvips]{graphicx}
\usepackage{amsbsy}
\usepackage{color}

\newcommand{\Her}{{\rm H}}
\newcommand{\La}{{\rm L}}
\newcommand{\sgn}{{\rm sgn}}

\newcommand{\rmd}{{\rm d}}
\newcommand{\rme}{{\rm e}}
\newcommand{\rmi}{{\rm i}}
\newcommand{\bra}{\langle}
\newcommand{\ket}{\rangle}
\newcommand{\mbfe}{\mathbf{e}}
\newcommand{\mbfj}{\mathbf{j}}

\newcommand{\mbfo}{\mathbf{o}}
\newcommand{\mbfp}{\mathbf{p}}

\newcommand{\mbfr}{\mathbf{r}}

\newcommand{\mbfv}{\mathbf{v}}

\newcommand{\mbfA}{\mathbf{A}}

\newcommand{\mbfE}{\mathbf{E}}

\newcommand{\mbfH}{\mathbf{H}}

\newcommand{\mbfrho}{\boldsymbol{\rho}}
\newcommand{\mbfsigma}{\boldsymbol{\sigma}}
\newcommand{\elf}{{\cal E}_\perp}
\newcommand{\Elf}{{\cal E}}
\newcommand{\ELF}{{\boldsymbol{\cal E}}}
\newcommand{\mgf}{{\cal B}}
\newcommand{\MGF}{{\boldsymbol{\cal B}}}

\begin{document}
\begin{article}
\begin{opening}
\title{Propagation in crossed electric and magnetic fields:\\The quantum source approach}
\runningtitle{Propagation in crossed electric and magnetic fields}
\author{Tobias \surname{Kramer}\email{tkramer@ph.tum.de}}
\institute{Physik-Department T30,
Technische Universit\"at M\"unchen\\
James-Franck-Stra{\ss}e,
85747 Garching, Germany}
\author{Christian \surname{Bracher}\email{cbracher@fizz.phys.dal.ca}}
\institute{Department of Physics and Atmospheric Science, Dalhousie
University\\
Halifax, N.S.\ B3H~3J5, Canada}
\begin{abstract}
The propagation of electrons in static and uniform electromagnetic fields is a standard topic of classical electrodynamics. The Hamilton function is given by a quadratic polynomial in the positions and momenta. The corresponding quantum-mechanical problem has been analyzed in great detail and the eigenfunctions and time evolution operators are well-known. Surprisingly, the energy-dependent counterpart of the time-evolution operator, the Green function, is not easily accessible. However in many situations one is interested in the evolution of a system that started with emitted particles that carry a specific energy. In the following we present a suitable approach to study this type of matter waves arising from a localized region in space. Two applications are discussed, the photodetachment current in external fields and the quantum Hall effect in a fermionic electron gas.
\end{abstract}
\keywords{Green function. Electric and magnetic fields. Hall effect.}
\end{opening}

\section{Introduction}

In quantum mechanics, static and uniform electric and magnetic fields are represented by a quadratic Hamiltonian (i.e.\ a second order polynomial in the canonical coordinates $r_i$ and momenta $p_i$). Moshinsky and Winternitz carried out a detailed group-theoretical analysis of this class of Hamiltonians and their eigenfunctions \cite{Moshinsky1980a}. Quadratic Hamiltonians are connected to linear canonical transformations \cite{Moshinsky1971a} and therefore a general phase-space approach gives valuable information for their classification in different dimensions. Nieto used the Moyal phase-space representation to develop a general method for deriving the time-evolution operator for quadratic Hamiltonians \cite{Nieto1992a}. However, its energy-dependent counterpart, the Green function, withstands such a systematic analysis and is not available in analytic form for many physical relevant potentials. Also other methods, like the Feynman path-integral approach, are not capable to derive the exact energy-dependent Green function.

In experiments, the energy of particles is often easier controlled than the time of travel.  Under these circumstances, the energy-dependent Green function is relevant for the description of the system.  Monochromatic particle sources arise in numerous applications of quantum mechanics. In accelerator physics sources located far away from the scattering region lead to boundary conditions in the form of incoming plane waves.  In this contribution, we study the behaviour of spatially localized electron sources in perpendicular, homogeneous electric and magnetic fields.  Our discussion is based on the framework of quantum source theory, a variant of the scattering formalism that is especially suited to describe scattering events restricted to a region of finite volume \cite{Bracher1999a,Bracher2003a,Kramer2002a,Kramer2003b}.  The idea of quantum sources was first promoted by Schwinger \cite{Schwinger1973a} but has not found widespread attention.  Therefore, we briefly introduce the concept and some basic results derived from it, and stress its connection to the propagator approach to quantum mechanics \cite{Feynman1965a}.  In fact, stationary elastic scattering at pointlike sources is fully described in terms of the energy Green function.  Several fundamental properties of the quantum system, like the scattering wave function, current density distribution, cross section, and local density of states, immediately follow from this functional.  Here, we explore in detail isotropic point sources in crossed external static fields both in two- and three-dimensional configuration space.  The results are in agreement with experimental findings in a recent photodetachment experiment, and offer an alternative viewpoint towards the anomalous Hall effect observed in low-dimensional semiconductor devices.

\section{Elastic scattering and quantum sources}
\label{sec:SourceForm}

In preparation for our later discussion, we illustrate the quantum source formalism using potential scattering in external static fields as an example.  We assume that the potential $V(\mathbf r)$ represents a localized disturbance, and that the charged quantum particles otherwise move in the external electromagnetic potentials $\mathbf A(\mathbf r)$, $\Phi(\mathbf r)$:
\begin{equation}
\label{eq:th01}
\mbfH = \mathbf H_0 + V(\mathbf r) = \frac1{2m} \left(\mbfp-q\mbfA(\mbfr)/c\right)^2 + q\Phi(\mathbf r)+ V(\mathbf r)\;.
\end{equation}
The scattering solutions $\psi(\mathbf r)$, which are eigenfunctions of the Hamiltonian $\mathbf H$ with energy $E$, then usually are decomposed into two parts, $\psi(\mathbf r) = \psi_{\rm in}(\mathbf r) + \psi_{\rm sc}(\mathbf r)$, where the incident wave is a solution for the unperturbed system $\mathbf H_0$:  $\mathbf H_0 \psi_{\rm in}(\mathbf r) = E\psi_{\rm in}(\mathbf r)$, whereas the remainder $\psi_{\rm sc}(\mathbf r)$ represents the scattering wave.  By comparison with (\ref{eq:th01}), we find that $\psi_{\rm sc}(\mathbf r)$ obeys:
\begin{equation}
\label{eq:th02}
\left[ E - \mbfH_0 - V(\mbfr)\right] \psi_{\rm sc}(\mathbf r) = V(\mathbf r)\psi_{\rm in}(\mathbf r) \;.
\end{equation}
Hence, $\psi_{\rm sc}(\mathbf r)$ is a solution to the inhomogeneous Schr\"odinger equation of the full Hamiltonian $\mbfH = \mathbf H_0 + V(\mathbf r)$, where we denote the right-hand side in (\ref{eq:th02}) as the source term $\sigma(\mathbf r)$:
\begin{equation}
\label{eq:SourceTerm}
\sigma(\mbfr):=V(\mbfr)\;\psi_{\rm in}(\mbfr) \;.
\end{equation}
Equation (\ref{eq:th02}) suggests the following physical interpretation: The incoming wave $\psi_{\rm in}(\mathbf r)$, via the perturbation $V(\mathbf r)$, feeds particles into the scattering wave $\psi_{\rm sc}(\mathbf r)$ that is governed by the Hamiltonian $\mbfH$.  While not commonly seen in standard quantum theory, inhomogeneous partial differential equations are familiar from other branches of physics, the heat conduction equation and Maxwell's equations being examples for the introduction of sources.  For these problems, a sophisticated mathematical framework in the form of Green functions has been developed.  Accordingly, we introduce the energy-dependent Green function $G(\mathbf r,\mathbf r';E)$ for the Hamiltonian $\mbfH$ defined via \cite{Economou1983a}
\begin{equation}
\label{eq:Multi1.3}
\left[ E - \mbfH_0 - V(\mathbf r) \right] G(\mathbf r,\mathbf r';E) = \delta(\mathbf  r - \mathbf r').
\end{equation}
Formally, the solution to equation~(\ref{eq:th02}) is given by a convolution integral comprising the source term and the Green function
\begin{equation}
\label{eq:Multi1.4}
\psi_{\rm sc}(\mathbf r) = \int {\rm d}^3r'\, G(\mathbf r,\mathbf r';E) \sigma(\mathbf r').
\end{equation}
We infer that the scattering wave generated by the source $\sigma(\mathbf r)$ allows for an interpretation as the linear superposition of ``fundamental'' waves $G(\mathbf r,\mathbf r';E)$ emitted from point sources $C\delta(\mathbf r-\mathbf r')$ located at $\mathbf r'$.

\subsection{Connection to the propagator}
\label{sec:Sources1}

In the continuous spectrum of $\mathbf H$, the Green function is not uniquely defined.  Depending on our choice for $G(\mathbf r,\mathbf r';E)$, we obtain a set of wave functions $\psi_{{\rm sc}}(\mbfr)$ that differ only by eigenfunctions $\psi_{\rm hom}(\mathbf r)$ of $\mbfH$.  This ambiguity is resolved by the demand that $G(\mathbf r,\mathbf r';E)$ presents a retarded solution that enforces outgoing-wave behaviour of the scattering wave $\psi_{\rm sc}(\mathbf r)$ at large distances from the source.  The representation of $G(\mathbf r,\mathbf r';E)$ as a Laplace transform of the quantum propagator $K(\mathbf r,t|\mathbf r',t_0)$ \cite{Feynman1965a} guarantees the proper choice of boundary conditions for the Green function \cite{Economou1983a}:
\begin{equation}
\label{eq:GEnergyIntegral}
G(\mbfr,\mbfr';E)
=-\frac{\rmi}{\hbar}\lim_{\eta\rightarrow 0_+}
\int_0^\infty \rmd T\;\rme^{\rmi E T/\hbar-\eta T/\hbar}\;K(\mbfr,T|\mbfr',0).
\end{equation}
where $K(\mbfr,t|\mbfr',t_0)$ denotes the coordinate space representation of the time evolution operator $\mathbf U(t,t_0)$
\begin{equation}\label{eq:PropagatorKernel}
K(\mbfr,t|\mbfr',t_0):=\bra\mbfr|\mathbf U(t,t_0)|\mbfr'\ket.
\end{equation}
Since for a conservative system, $\mathbf U(T,0) = \exp(-\rmi \mathbf H T/\hbar)$ holds, we may formally integrate (\ref{eq:GEnergyIntegral}) to obtain:
\begin{equation}\label{eq:GreenEnergy2}
G(\mbfr,\mbfr';E)=\lim_{\eta\rightarrow 0_+}
\left\bra \mbfr \left| \frac{1}{E-\mbfH+\rmi\eta} \right| \mbfr' \right\ket.
\end{equation}
Therefore, the Green function represents the resolvent assigned to the Hamiltonian $\mathbf H$ in configuration space.  According to (\ref{eq:GreenEnergy2}), $G(\mbfr,\mbfr';E)$ indeed acts as an ``inverse'' to the operator $E-\mathbf H$.

At least in principle, knowledge of the full Green function permits the exact evaluation of the scattering wave $\psi_{\rm sc}(\mathbf r)$ (\ref{eq:Multi1.4}).  In general, however, $G(\mathbf r,\mathbf r';E)$ is not available in analytic form.  In the favourable situation that we can find an expression for the Green function $G_0(\mathbf r,\mathbf r';E)$ associated with the unperturbed Hamiltonian $\mathbf H_0$, $G(\mathbf r,\mathbf r';E)$ formally may be expanded into a series via the Dyson equation:
\begin{equation}
\label{eq:Dyson}
\frac1{E-\mathbf H} = \frac1{E-\mathbf H_0} \left[ 1 + \mathbf V \frac1{E-\mathbf H} \right].
\end{equation}
Replacing $G(\mathbf r,\mathbf r';E)$ by $G_0(\mathbf r,\mathbf r';E)$, i.~e., neglect of the rescattering terms that involve the perturbation $\mathbf V$, is equivalent to the leading order of perturbation theory in the conventional scattering formalism, which we will endorse in the following.

While the quantum propagators $K(\mathbf r,t|\mathbf r',t_0)$ are tabulated for a fairly extensive set of potentials \cite{Grosche1998a,Kleber1994a}, few energy Green functions are available in closed form for problems in more than one spatial dimension.  This list includes free particles in two and three dimensions, as well as the Coulomb problem \cite{Hostler1963a,Hostler1964a}, uniformly accelerated quantum motion \cite{Bracher1998a,Dalidchik1976a,Gottlieb1991a,Slonim1976a}, the isotropic harmonic oscillator \cite{Bakhrakh1971a}, motion in a homogeneous magnetic field \cite{Dodonov1975a,Gountaroulis1972a}, and parallel electric and magnetic fields \cite{Fabrikant1991a,Kramer2001a}, all in three-dimensional configuration space.

\subsection{Currents generated by quantum sources}
\label{sec:Sources2}

A first obvious quantity of interest are the currents associated with the scattering wave $\psi_{\rm sc}(\mathbf r)$ generated by the source $\sigma(\mathbf r)$ (\ref{eq:Multi1.4}).  The current density distribution $\mathbf j(\mathbf r)$ is defined in the usual fashion via
\begin{equation}
\label{eq:currdens}
\mathbf j(\mathbf r) = \frac{\hbar}{m} \Im[\psi_{\rm sc}(\mathbf r)^* \boldsymbol\nabla \psi_{\rm sc}(\mathbf r)]-\frac{q \mathbf{A}(\mbfr)}{m}{|\psi_{\rm sc}(\mbfr)|}^2,
\end{equation}
where $\mathbf A(\mathbf r)$ denotes the vector potential, and displays the spatial distribution of the quanta in the scattering wave, i.~e., is directly related to the differential cross section of the scattering process.  Integration of $\mathbf j(\mathbf r)$ over a surface enclosing $\sigma(\mathbf r)$ will yield the total current $J(E)$ emitted by the source which, in turn, is a measure of the total scattering rate.  For a concise expression, we first note that the inhomogeneous Schr\"odinger equation (\ref{eq:th02}) gives rise to a modified equation of continuity.  Instead of $\nabla\cdot\mbfj(\mbfr)=0$, valid for a stationary system in the absence of sources, we now find:
\begin{equation}
\label{eq:Multi1.6}
\boldsymbol\nabla\cdot\mathbf j(\mathbf r) = - \frac2\hbar \Im\left[ \sigma(\mathbf r)^* \psi_{\rm sc}(\mathbf r) \right].
\end{equation}
Thus, the inhomogeneity $\sigma(\mathbf r)$ acts also as a source for the particle current $\mathbf j(\mathbf r)$.  Since the current is conserved outside the source region, the surface integral may be replaced by a spatial integration over $\boldsymbol\nabla\cdot\mathbf j(\mathbf r)$ covering the source volume, and upon insertion of (\ref{eq:Multi1.4}) for the scattering wave in (\ref{eq:Multi1.6}), we obtain a bilinear expression for the total flux $J(E)$:
\begin{equation}
\label{eq:CurrentTotal}
J(E) = - \frac2\hbar \Im\left[ \int {\rm d}^3r \int {\rm d}^3r' \sigma(\mathbf r)^* G(\mathbf r,\mathbf r';E) \sigma(\mathbf r') \right] \;.
\end{equation}

For pointlike sources $\sigma(\mbfr)=C \delta(\mbfr-\mbfr')$, where $C$ is a measure for the source strength, the calculation of the scattering currents simplifies considerably.  In this case, the scattering wave is a multiple of the Green function, $\psi_{\rm sc}(\mathbf r) = C G(\mathbf r,\mathbf r';E)$, and the pattern of the current distribution follows from (\ref{eq:currdens}).  Point sources yield a particularly simple expression for the total cross section:
\begin{equation}
\label{eq:JDelta}
J(\mathbf r';E) = - \frac{2{|C|}^2}{\hbar} \lim_{\mathbf r\rightarrow \mathbf r'}
\Im\left\{G(\mathbf r,\mathbf r';E)\right\}.
\end{equation}
In passing, we remark that for $\mathbf r\rightarrow \mathbf r'$, the Green function $G(\mathbf r,\mathbf r';E)$ diverges in more than one spatial dimension, while its imaginary part remains well-defined in the limit and is proportional to the scattering rate.  The statement (\ref{eq:JDelta}) is closely related to the optical theorem of conventional scattering theory \cite{Sakurai1994a}.

For reference, we list the total currents emitted by a free-particle point source of unit strength ($C=1$) in one-, two-, and three-di\-men\-sio\-nal configuration space.  For $E>0$, they read:
\begin{equation}
\label{eq:JFree}
J_{\rm free}^{\rm (1D)}(E) = \frac{2m}{\hbar^3 k} \;,\qquad
J_{\rm free}^{\rm (2D)}(E) = \frac{m}{\hbar^3}    \;,\qquad
J_{\rm free}^{\rm (3D)}(E) = \frac{m k}{\pi \hbar^3} \;,
\end{equation}
where $k =\sqrt{2mE}/\hbar$ denotes the wave number of the particles.

\subsection{Density of States}
\label{sec:Sources3}

Somewhat surprisingly, the local density of states (LDOS) $n(\mbfr';E)$ of the quantum system, i.~e., the accumulated density $|\psi(\mathbf r)|^2$ of the eigenfunctions of the system with energy $E$, evaluated at $\mathbf r'$, is, apart from a prefactor, identical to the total current $J(\mathbf r';E)$ emitted by a point source located at the same position.  Formally, this equivalence is established from equation (\ref{eq:GreenEnergy2}) by setting $\mathbf r=\mathbf r'$ and using the distribution relation \cite{Halperin1952a} $\Im[(z + \rmi\eta)^{-1}] = -\pi\,\sgn\,\eta\cdot \delta(z)$ that holds in the limit $\eta\rightarrow 0$.  Thus, we formally obtain:
\begin{equation}
\label{eq:th31}
\Im[G(\mathbf r',\mathbf r';E)] = -\pi \left\langle \mathbf r' \left| \delta(E-\mathbf H) \right|\mathbf r'\right\rangle \;.
\end{equation}
The right-hand side of this relation formally contains the spatial representation of the density of states operator $\delta(E-\mathbf H)$, and we conclude that the LDOS is linked to the imaginary part of the Green function.  In conjunction with (\ref{eq:JDelta}), this implies that the density of states is directly proportional to the previously defined total current $J(\mathbf r';E)$:
\begin{equation}
\label{eq:JDOS}
J(\mbfr';E) = \frac{2\pi}{\hbar} {|C|}^2 n(\mbfr';E).
\end{equation}
We note here that the localized eigenstates of $\mathbf H$ that make up the discrete spectrum of the Hamiltonian are irrelevant for the imaginary part of the Green function, and thus do not contribute to the current.  It is the unbounded solutions in the continuous spectrum of $\mathbf H$ that are entirely responsible for the outgoing wave character of $G(\mathbf r,\mathbf r';E)$ and constitute $J(\mathbf r';E)$.  From (\ref{eq:JDOS}), we conclude that a non-vanishing density of states is therefore directly related to an extended flow pattern in position space. We will show examples of this behaviour in Section~\ref{sec:FlowPattern}.

\subsection{Construction of the Green function}
\label{sec:Sources4}

In the following, we will briefly discuss some techniques that are useful in establishing the energy Green function for simple systems.

\paragraph*{Matching of solutions}

For one-dimensional systems, the inhomogeneous Schr{\"o}dinger equation (\ref{eq:th02}) reduces to a linear ordinary differential equation of second order, and if it exists at all, the Green function $G(z,z';\epsilon)$ is always available as a product of solutions $\psi_{\epsilon,<}(z_<)$ and $\psi_{\epsilon,>}(z_>)$ that behave regularly in the sectors $z\rightarrow \pm\infty$, respectively, and are matched at the source position $z=z'$:
\begin{equation}
\label{eq:th41}
G(z,z';\epsilon) = \frac{2m}{\hbar^2} \frac{\psi_{\epsilon,<}(z_<)\psi_{\epsilon,>}(z_>)}{W[\psi_{\epsilon,<},\psi_{\epsilon,>}]} \;.
\end{equation}
Here we introduced the symbols $z_< = \min(z,z')$ and $z_> = \max(z,z')$, and $W[\psi_{\epsilon,<},\psi_{\epsilon,>}]$ denotes the Wronskian of the two solutions.  The basic example for this strategy is the free particle problem in one spatial dimension, where for $E>0$ ($E<0$) $\psi_{\epsilon,<}(z_<)$ and $\psi_{\epsilon,>}(z_>)$ are outgoing (evanescent) waves in either direction:
\begin{equation}
\label{eq:th42}
G_{\rm free}^{(1D)}(z,z';E) =
\left\{
\begin{array}{cc}
-\frac m{\hbar^2\kappa} \exp(-\kappa|z-z'|) & (E<0) \;, \\
&\\
-\frac{\rmi m}{\hbar^2k} \exp(\rmi k|z-z'|) & (E>0) \;.
\end{array}\right.
\end{equation}
Here, $k=\sqrt{2mE}/\hbar$, $\kappa = \sqrt{-2mE}/\hbar$ again denote the wave number of the particle.  In passing, we remark that the few higher-dimensional Green functions that can be found in analytical form usually have been determined by formal extensions of this technique \cite{Bakhrakh1971a,Bracher1998a,Hostler1963a,Slonim1976a}.

\paragraph*{Eigenfunction expansion}

In section~\ref{sec:Sources1}, we found a formal position space representation for the retarded energy Green function as a special resolvent of the Hamiltonian operator $\mathbf H$.  Expanding (\ref{eq:GreenEnergy2}) into a complete set of eigenstates $|\psi_\epsilon\rangle$ of $\mathbf H$ (where $\mathbf H|\psi_\epsilon\rangle = \epsilon |\psi_\epsilon\rangle$), we may alternatively express the Green function as a sum over all properly normalized eigenfunctions $\psi_\epsilon(\mathbf r) = \langle \mathbf r|\psi_\epsilon\rangle$ of the system:
\begin{equation}
\label{eq:th43}
G(\mbfr,\mbfr';E)=\lim_{\eta\rightarrow 0_+} \sum_{|\psi_\epsilon\rangle}
\frac{\psi_\epsilon(\mathbf r')^* \psi_\epsilon(\mathbf r)}{E - \epsilon +\rmi\eta} \;.
\end{equation}
We will encounter an example of this decomposition in Section~\ref{sec:Green2DEigenfunction}.

\paragraph*{Complex convolution}

We noted before that the quantum propagator $K(\mathbf r, t|\mathbf r',t_0)$ is generally more easily available than the Green function $G(\mathbf r,\mathbf r';E)$ \cite{Grosche1998a,Kleber1994a}.   In part, this situation is the consequence of the simple composition properties of $K(\mathbf r, t|\mathbf r',t_0)$.  Assume that the (conservative) Hamiltonian $\mathbf H$ of the system can be written as the sum of commuting, lower-dimensional parts:  $\mathbf H=\mathbf H_1 + \mathbf H_2$, where $\mathbf H_1\mathbf H_2 = \mathbf H_2\mathbf H_1$.  Then, the corresponding evolution operator obeys $\mathbf U(T,0) = \rme^{\rmi\mathbf H T/\hbar} = \rme^{\rmi\mathbf H_1 T/\hbar}\rme^{\rmi\mathbf H_2 T/\hbar} = \mathbf U_1(T,0)\mathbf U_2(T,0)$, and thus reduces to a product of its constituents.  This property is transferred to their spatial representations, the propagators:
\begin{equation}
\label{eq:th44}
K(\mathbf r, t|\mathbf r',t_0) = K_1(\mathbf r_1, t|\mathbf r'_1,t_0) K_2(\mathbf r_2, t|\mathbf r'_2,t_0) \;,
\end{equation}
where $\mathbf r_1,\mathbf r_2$ are the projections of $\mathbf r$ onto the subspaces of $\mathbf H_1$, $\mathbf H_2$.

Unfortunately, the simple multiplicative property (\ref{eq:th44}) does not extend to the energy domain.  We may, however, exploit it to derive a corresponding statement for the Green functions $G_\nu(\mathbf r_\nu, \mathbf r'_\nu; E)$.  Equation~(\ref{eq:GEnergyIntegral}) shows that the evolution operator $\mathbf U(T,0)$ and the resolvent operator $[E-\mathbf H]^{-1}$, which yields the Green function in configuration space (\ref{eq:GreenEnergy2}), are linked through a Laplace transform.  Since the image of a product of Laplace transforms is represented by the convolution integral of the images of the factors, we obtain $[E-\mathbf H]^{-1} = \frac{\rmi}{2\pi} \int\rmd E'\, [E'-\mathbf H_1]^{-1}[E- E'-\mathbf H_2]^{-1}$, or, in  position representation:
\begin{equation}
\label{eq:ComplexConvolution}
G(\mbfr,\mbfr';E) = \frac{\rmi}{2\pi}\int_{-\infty}^{\infty}\rmd E'\;
G_1(\mathbf r_1, \mathbf r'_1; E') \; G_2(\mathbf r_2, \mathbf r'_2; E-E').
\end{equation}
Due to the generally complicated form of the energy-dependent Green function, the practical value of this relation is limited. We will, however, present an application in the following section.

\section{Matter waves in crossed electric and magnetic fields}

As our example of interest, we study quantum sources of charged particles in an environment of homogeneous, static electric and magnetic fields $\ELF$, $\MGF$.  The Hamiltonian $\mathbf H$ in this case may be written as the sum of commuting parts $\mathbf H_\parallel(r_\parallel)$ and $\mathbf H_\perp(\mathbf r_\perp)$ in the sense of Section~\ref{sec:Sources4} and reads:
\begin{eqnarray}
\label{eq:crossed01}
\mathbf H_\parallel(r_\parallel) &=& -\frac{\hbar^2}{2m} \frac{\partial^2}{\partial r_\parallel^2} - qr_\parallel\Elf_\parallel \;, \\
\label{eq:crossed02}
\mathbf H_\perp(\mathbf r_\perp) &=& \frac{1}{2m} \left[ -\rmi\hbar\boldsymbol\nabla_\perp - \frac q2 (\MGF \times \mathbf r_\perp) \right]^2 - q \mathbf r_\perp\cdot\mathbf\ELF_\perp \;.
\end{eqnarray}
Here, the subscripts in $r_\parallel$ and $\mathbf r_\perp$ denote the directions parallel and perpendicular to the magnetic field $\MGF$, respectively.  We chose the electromagnetic potentials $\mathbf A(\mathbf r) = \frac12 (\MGF \times \mathbf r)$ and $\Phi(\mathbf r) = - \mathbf r\cdot\ELF$ as particular gauge in $\mathbf H = \mathbf H_\parallel(r_\parallel) + \mathbf H_\perp(\mathbf r_\perp)$ (\ref{eq:crossed01}), (\ref{eq:crossed02}), but all observable quantities, e.~g.\ the currents $\mathbf j(\mathbf r)$ and $J(E)$ (\ref{eq:currdens}), (\ref{eq:CurrentTotal}), are invariant under gauge transformations, unlike the propagator and Green function.  (We note that under a change of gauge field $\chi(\mathbf r,t)$, the source term (\ref{eq:SourceTerm}) must be correspondingly modified.)

\subsection{The quantum propagator}
\label{sec:crossed1}

According to the composition properties outlined in Section~\ref{sec:Sources4}, the propagator for a particle in homogeneous fields $\ELF$, $\MGF$ at arbitrary angle may be written as a product (\ref{eq:th44}):
\begin{equation}
\label{eq:crossed11}
K_{\ELF,\MGF}(\mathbf r, t|\mathbf r',0) = K_\parallel(r_\parallel, t|r'_\parallel,0) K_\perp(\mathbf r_\perp, t|\mathbf r'_\perp,0) \;,
\end{equation}
Here, $K_\parallel(r_\parallel, t|r'_\parallel,0)$ is the propagator for a uniformly accelerated particle in one dimension that has been known from the beginnings of quantum mechanics \cite{Bracher1998a,Feynman1965a,Kennard1927a,Kleber1994a}:
\begin{eqnarray}
\label{eq:crossed12}
\lefteqn{K_\parallel(r_\parallel, t|r'_\parallel,0) = \sqrt{\frac m{2\pi\rmi\hbar t}}} \\\nonumber
&& \times\, \exp\left\{ \frac\rmi{\hbar} \left[ \frac m{2t}(r_\parallel - r'_\parallel)^2 + \frac{qt}2 \Elf_\parallel(r_\parallel + r'_\parallel) - \frac{q^2\Elf_\parallel^2t^3}{24 m} \right] \right\} \;,
\end{eqnarray}
whereas $K_\perp(\mathbf r_\perp, t|\mathbf r'_\perp,0)$, the propagator for a charge moving in two dimensions subject to perpendicular electric and magnetic fields, was unraveled much later \cite{Horing1986a,Nieto1992a,Souza1988a}:
\begin{eqnarray}
\label{eq:crossed13}
\lefteqn{
K_\perp(\mathbf r_\perp, t|\mathbf r'_\perp,0) = \frac{m\omega_L}{2\pi\rmi\hbar\sin(\omega_L t)}
\exp \left\{ \frac\rmi\hbar \left[ \frac q2 \MGF\cdot(\mathbf r'_\perp \times \mathbf r_\perp) \right.\right.}\\\nonumber
&&
+ \frac{qt}2 \ELF_\perp\cdot(\mathbf r_\perp + \mathbf r'_\perp)
+ 2m \mathbf v_D \cdot (\mathbf r_\perp - \mathbf r'_\perp) - \frac m2 v_D^2 t \\\nonumber
&& \left.\left.
+ \frac{m\omega_L}2 \cot(\omega_L t) \left[ (\mathbf r_\perp - \mathbf r'_\perp)^2 - 4 \mathbf v_D \cdot (\mathbf r_\perp - \mathbf r'_\perp) t + v_D^2 t^2 \right] \right] \right\} \;.
\end{eqnarray}
In this expression, we introduced the Larmor frequency $\omega_L$ and the drift velocity $\mathbf v_D$:
\begin{equation}
\label{eq:crossed14}
\omega_L = q\mgf/(2m) \;, \qquad \mathbf v_D = (\ELF \times \MGF) / \mgf^2 \;.
\end{equation}
Interestingly, Schwinger derived the relativistic propagator much earlier \cite{Schwinger1951a}, but apparently no simple transition to the non-relativistic case exists.

The equations~(\ref{eq:crossed12}) and (\ref{eq:crossed13}) reveal a simple symmetry property of the propagator under translations of the coordinate origin:
\begin{equation}
\label{eq:crossed15}
K_{\ELF,\MGF}(\mathbf r, t|\mathbf r',0) = \exp\left\{ \frac{\rmi q}\hbar \left[ \frac12 \MGF\cdot(\mathbf r'\times \mathbf r) + \mathbf r'\cdot\ELF t \right] \right\} K_{\ELF,\MGF}(\mathbf r - \mathbf r', t|\mathbf o,0) \;.
\end{equation}
(Alternatively, the symmetry (\ref{eq:crossed15}) may be viewed as the effect of a gauge transform that shifts the origin of the potentials \cite{Kramer2003b}.)  Of primary interest in our study is the corresponding energy Green function $G_{\ELF,\MGF}(\mathbf r,\mathbf r';E)$ (\ref{eq:Multi1.3}).  Its analytical expression is unknown, however.  Hence, the results displayed in subsequent figures were obtained by numerical evaluation of the integral representation (\ref{eq:GEnergyIntegral}).  Fortunately, the symmetry (\ref{eq:crossed15}) of the propagator, together with (\ref{eq:GEnergyIntegral}), permits to predict the behaviour of $G_{\ELF,\MGF}(\mathbf r,\mathbf r';E)$ under coordinate transformations:
\begin{equation}
\label{eq:EBGauge}
G_{\ELF,\MGF}(\mbfr,\mbfr';E) = \exp\left\{\frac{\rmi q}{2\hbar}\MGF\cdot(\mbfr'\times\mbfr)\right\}
G_{\ELF,\MGF}(\mbfr-\mbfr',\mbfo;E + q\mbfr'\cdot\ELF) \;.
\end{equation}
This relation immediately extends any result obtained for $\mathbf r' = \mathbf o$ to general source locations $\mathbf r' \neq \mathbf o$.

\subsection{Purely magnetic field}
\label{sec:From2Dto3D}

As a simple example, we first inquire into the dynamics of a charge in a purely magnetic field ($\ELF =\mathbf o$).  In two spatial dimensions, the relevant propagator $K_\perp(\mathbf r_\perp, t|\mathbf o,0)$ (\ref{eq:crossed13}) reduces to:
\begin{equation}
\label{eq:KB}
K_\MGF^{(2D)}(\mathbf r_\perp,t|\mbfo,0) = \frac{m\omega_L}{2\pi\rmi\hbar\sin(\omega_L t)}
\exp\left\{\frac{\rmi m\omega_L}{2\hbar}\mathbf r_\perp^2 \cot\left(\omega_L t\right)\right\} \;.
\end{equation}
As the propagator is periodic in $t$, we may expand it into a series using the generating function of the Laguerre polynomials $\La_k(z)$ \cite{Abramowitz1965a}:
\begin{equation}
\label{eq:LaguerreTheorem}
\frac1{1-z}\,\exp\left\{ \frac{xz}{z-1} \right\} = \sum_{n = 0}^\infty \La_n(x)\,z^n \;,
\end{equation}
where we set $z = \rme^{2\rmi\omega_L t}$ and $x = m\omega_L r_\perp^2/\hbar$.  This procedure yields:
\begin{equation}
\label{eq:crossed21}
K_\MGF^{(2D)}(\mathbf r_\perp,t|\mbfo,0) = \frac{m\omega_L}{\pi\hbar} \rme^{- m\omega_L r_\perp^2/2\hbar}  \sum_{n = 0}^\infty \La_n\left( \frac{m\omega_L r_\perp^2}\hbar \right) \rme^{-\rmi (2n+1) \omega_L t} \;.
\end{equation}
This form is easily recognized as the decomposition of the time evolution operator $\mathbf U_\perp(t,0)$ into the eigenfunctions of $\mathbf H_\perp$ (\ref{eq:crossed02}) populating the Landau levels $E_n = (2n+1)\hbar\omega_L$.

\begin{figure}[t]
\centerline{
\includegraphics[width=0.5\textwidth]{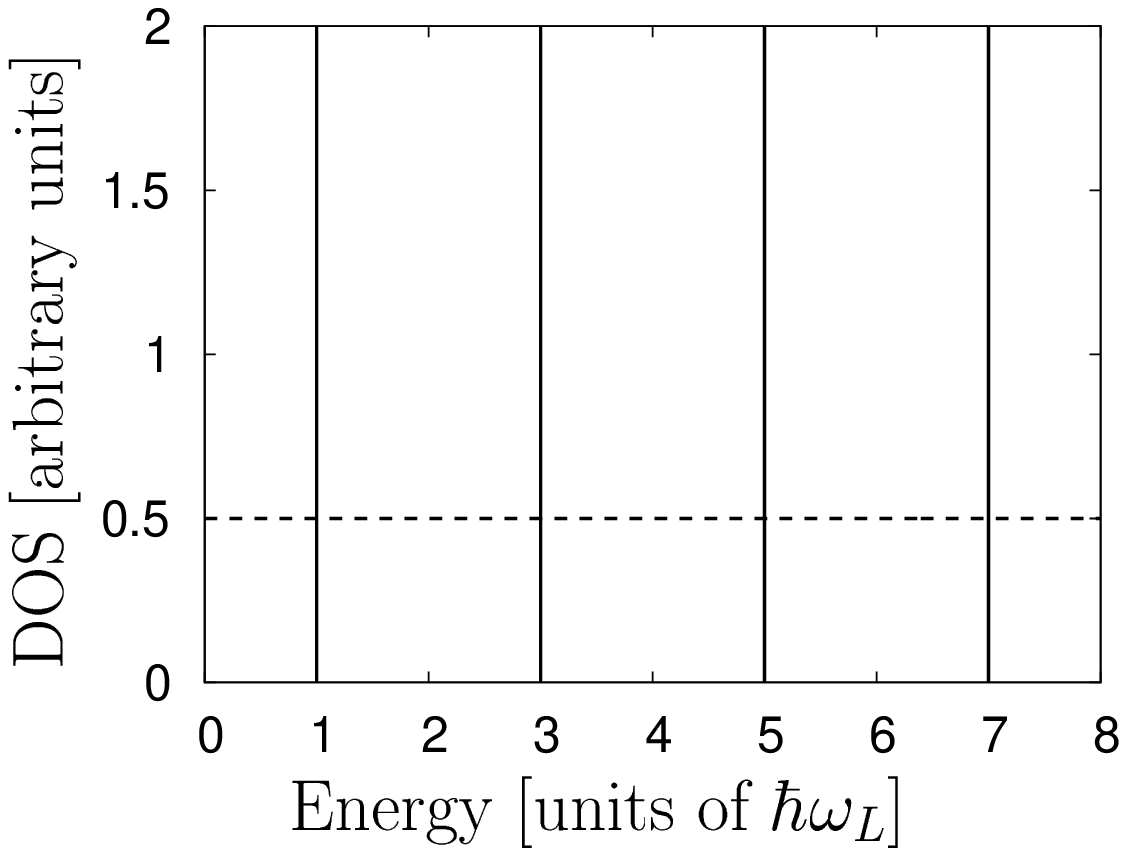}
\includegraphics[width=0.5\textwidth]{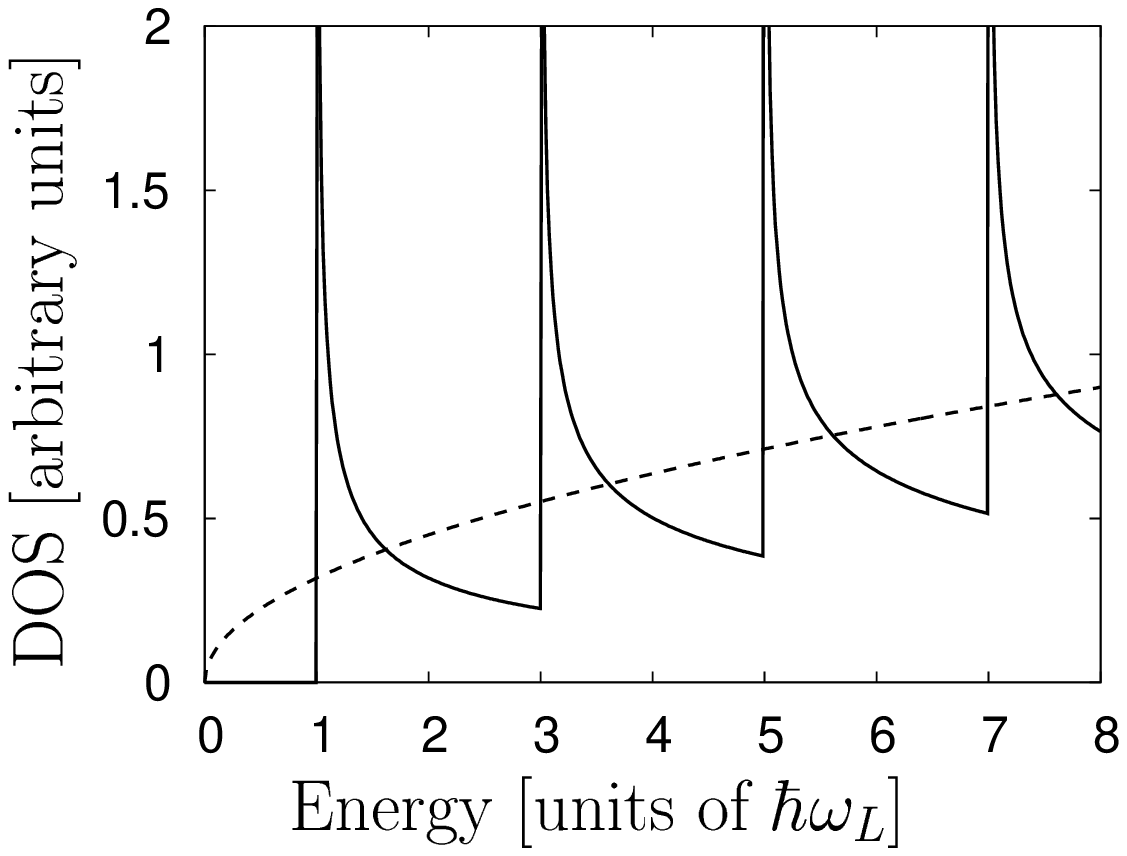}
}
\caption[Density of states in an electron gas with magnetic field]
{Electronic density of states in two (left panel) and three dimensions (right panel) in a purely magnetic field. The dashed line denotes Wigner's threshold law (\ref{eq:JFree}), valid for free particles.}\label{fig:DOSB}
\end{figure}

The two-dimensional energy Green function $G_\MGF^{(2D)}(\mathbf r_\perp,\mbfo;E)$ follows from (\ref{eq:crossed21}) after the Laplace transform (\ref{eq:GEnergyIntegral}) which immediately yields an infinite series expression:
\begin{equation}
\label{eq:GreenB2D}
G_{\MGF}^{(2D)}(\mathbf r_\perp,\mbfo;E) = \frac{m\omega_L}{\pi\hbar} \rme^{- m\omega_L r_\perp^2/2\hbar}  \lim_{\eta\rightarrow 0} \sum_{n=0}^{\infty}
\frac{\La_n\left( m\omega_L r_\perp^2 /\hbar \right)}{E-\hbar\omega_L (2n+1)+\rmi\eta} \;,
\end{equation}
which clearly resembles the formal eigenfunction expansion (\ref{eq:th43}).  In the important case $\mathbf r_\perp \rightarrow \mathbf o$, we extract the density of states for a two-dimensional gas of charges subject to a magnetic field:
\begin{equation}
\label{eq:JB2D}
n_{\MGF}^{(2D)}(E)=-\frac{1}{\pi}\Im\{G_{\MGF}^{(2D)}(\mbfo,\mbfo,E)\}
=\frac{m\omega_L}{\pi\hbar}\sum_{n=0}^\infty \delta\bigl[E-\hbar\omega_L(2n+1)\bigr] \;.
\end{equation}
Here, we again made use of the distribution relation $\Im[(z + \rmi\eta)^{-1}] = -\pi\,\sgn\,\eta\cdot \delta(z)$ \cite{Halperin1952a} that holds for $\eta \rightarrow 0$.  The resulting discrete $\delta$--array is indicated in Figure~\ref{fig:DOSB}.  Equation~(\ref{eq:JB2D}) expresses the fact that the eigenstates take on only the discrete energy values $E_n$.

In three spatial dimensions, we have to multiply the propagator $K_\MGF^{(2D)}(\mathbf r_\perp,t|\mbfo,0)$ with the free-particle propagator in one dimension $K_{\rm free}^{(1D)}(r_\parallel,t|0,0) = \sqrt{m/(2\pi\rmi\hbar t)} \exp\bigl[ \rmi m r_\parallel^2/(2\hbar t) \bigr]$ that follows from (\ref{eq:crossed12}) once we set $\Elf_\parallel = 0$.  Its Laplace transform (\ref{eq:GEnergyIntegral}) is the free-particle energy Green function $G_{\rm free}^{(1D)}(r_\parallel,0;E)$ (\ref{eq:th42}) that we derived in the preceding section.  Similarly, we may transform the product of $K_{\rm free}^{(1D)}(r_\parallel,t|0,0)$ with the series expansion of $K_\MGF^{(2D)}(\mathbf r_\perp,t|\mbfo,0)$ (\ref{eq:crossed21}) to determine the three-dimensional Green function of a charge in a homogeneous magnetic field:
\begin{eqnarray}
\label{eq:crossed22}
\lefteqn{G_{\MGF}^{(3D)}(\mathbf r,\mathbf o;E) = \frac{m\omega_L}{\pi\hbar} \,\rme^{- m\omega_L r_\perp^2/2\hbar}} \\\nonumber
&& \times\, \sum_{n=0}^{\infty} \La_n\left( m\omega_L r_\perp^2 /\hbar \right)
G_{\rm free}^{(1D)}\bigl[r_\parallel,0;E - (2n+1)\hbar\omega_L\bigr] \;.
\end{eqnarray}
(A representation in closed form is stated in \cite{Dodonov1975a}.)  Only the terms with positive effective energy ($E - (2n+1)\hbar\omega_L > 0$) contribute to the density of states $n_{\MGF}^{(3D)}(E) = -\frac1\pi \Im[G_{\MGF}^{(3D)}(\mathbf o,\mathbf o;E)]$, as comparison with (\ref{eq:th42}) shows.  With the help of (\ref{eq:JFree}), (\ref{eq:th42}) we find:
\begin{equation}
\label{eq:crossed23}
n_{\MGF}^{(3D)}(E) = \frac{m^{3/2}\omega_L}{\sqrt{2}\,\pi^2\hbar^2}
\sum_{n=0}^\infty \frac{\Theta\bigl[E-(2n+1)\hbar\omega_L\bigr]}{\sqrt{E-(2n+1)\hbar\omega_L}}.
\end{equation}
This superposition of effectively one-dimensional free-particle sources is displayed in Figure~\ref{fig:DOSB}.  In passing, we point out that the limes $\mgf\rightarrow 0$ in (\ref{eq:crossed23}) is not well-defined; only after averaging over a small energy range $\delta E$, the Wigner free-particle law $n_{\rm free}^{(3D)}(E) = mk/2\pi^2\hbar^2$ (\ref{eq:JFree}) will emerge.  Finally, we note that the result (\ref{eq:crossed22}) extends to the case of parallel electric and magnetic fields $\ELF\parallel\MGF$, once the free particle Green function in the sum is replaced by the one-dimensional Green function for a uniformly accelerated particle \cite{Fabrikant1991a,Kramer2001a}.

\subsection{Crossed electric and magnetic fields}
\label{sec:CrossedFields}

Unlike the case of a purely magnetic or parallel fields, the energy-dependent Green function $G_{\ELF,\MGF}(\mathbf r,\mathbf o;E)$ for a particle in crossed electric and magnetic fields $\ELF$, $\MGF$ is not available in closed form.  Thus, in this section we limit our considerations to the current emitted by a point source, or equivalently, the density of states $n(\mathbf o;E)$.  In a purely magnetic field, the degeneracy of the energy spectrum leads to peculiar shapes of the density of states functionals $n_{\MGF}(E)$ (see Figure~\ref{fig:DOSB}).  The presence of an additional perpendicular electric field $\ELF$ lifts these degeneracies and renders a broadened Landau level structure.

\subsubsection{Density of states in two dimensions}
\label{subsec:DOS2D}

We first examine the two-dimensional case.  Here, the Green function $G_{\ELF\times\MGF}^{(2D)}(\mathbf r_\perp,\mathbf o;E)$ in perpendicular fields is formally given by the Laplace transform (\ref{eq:GEnergyIntegral}) of the propagator $K_\perp(\mathbf r_\perp, t|\mathbf o,0)$ (\ref{eq:crossed13}).  Since we are only interested in the imaginary part of the Green function at $\mathbf r_\perp =\mathbf o$ (\ref{eq:JDelta}), (\ref{eq:JDOS}) we may rewrite this relation and express the density of states $n_{\ELF\times\MGF}^{(2D)}(\mathbf o;E)$ as the Fourier transform of the propagator:
\begin{equation}
\label{eq:crossed31}
n_{\ELF\times\MGF}^{(2D)}(\mathbf o;E) = \frac1{2\pi\hbar} \int_{-\infty}^\infty \rmd T\, \rme^{\rmi ET/\hbar}\, K_\perp(\mathbf o, T|\mathbf o,0) \;.
\end{equation}
(Formally, the density of states operator $\delta(E-\mathbf H)$ (\ref{eq:th31}) is the Fourier transform of the time evolution operator $\mathbf U(T,0) = \exp(-\rmi \mathbf H T/\hbar)$, and the identity (\ref{eq:crossed31}) follows in configuration space representation.)  Alternatively, we may determine $n_{\ELF\times\MGF}^{(2D)}(\mathbf o;E)$ by direct summation over a complete set of eigenstates of $\mathbf H_\perp(\mathbf r_\perp)$.  We will explore both routes below.

\paragraph*{Eigenfunction method}
\label{sec:Green2DEigenfunction}

A complete set of eigenfunctions for a charge in perpendicular fields is conveniently determined in the Landau gauge $\mathbf A = (-\mgf y,0,0)$ \cite{Johnson1983a}. Here, we assume that the magnetic field points into the $z$ direction while the electric field component $\ELF_\perp$ is aligned to the $y$--axis, so the charges drift along the $x$--axis. The corresponding Hamiltonian $\mathbf H'_\perp(x,y)$ is given by:
\begin{equation}
\label{eq:HEB}
\mathbf H'_\perp(x,y) = \frac 1{2m} \left( -\rmi\hbar\partial_x + q\mgf y \right)^2 - \frac 1{2m} \partial_y^2 - q\elf y \;.
\end{equation}
The eigenfunctions are products of shifted oscillator functions with a plane wave in drift direction, and read properly normalized:
\begin{equation}
\label{eq:HEBEigenfunction}
\psi_{n,y_c}(x,y) = \left( \frac{q\mgf}{2\pi\hbar} \right)^{1/2} \exp\left[ \frac\rmi\hbar \left( mv_D - q\mgf y_c \right) x \right] \; \frac1{\sqrt l} \,u_n\left( \frac{y-y_c}l \right) \;,
\end{equation}
where
\begin{equation}
\label{eq:QHO}
u_n(\xi) = \left( \frac1{2^n n!\,\sqrt\pi} \right)^{1/2} \, \rme^{-\xi^2/2} \Her_n\left(\xi\right) \;.
\end{equation}
Here, the magnetic length $l = \sqrt{\hbar/(q\mgf)}$ determines the extension of the wave function in the direction of $\ELF_\perp$, while the continuous variable $y_c$ denotes its centroid. $\Her_n\left(\xi\right)$ is a Hermite polynomial of order~$n$ \cite{Abramowitz1965a}.  The terms in the corresponding eigenenergy $E_n(y_c)$:
\begin{equation}
\label{eq:crossed32}
E_n(y_c) = (2n+1)\hbar\omega_L + mv_D^2/2 - q\elf y_c \;,
\end{equation}
reflect the Landau level, the kinetic energy of the drift motion and the potential energy in the electric field, respectively.  Summation over all eigenstates (\ref{eq:HEBEigenfunction}) yields the density of states (\ref{eq:crossed31}):
\begin{eqnarray}
\label{eq:crossed33}
n_{\ELF\times\MGF}^{(2D)}(\mathbf o;E) & = &\sum_{n=0}^\infty \int_{-\infty}^\infty \rmd y_c\, \delta\bigl[ E - E_n(y_c) \bigr] \, \bigl| \psi_{n,y_c}(\mathbf o) \bigr|^2 \\\nonumber
& = &
\frac{qB}{2\pi\hbar} \sum_{n=0}^\infty \frac 1{2^n n! \sqrt\pi\,\Gamma}\, \rme^{-E_n^2/\Gamma^2}\,\bigl[ \Her_n(E_n/\Gamma )\bigr]^2 \;.
\end{eqnarray}
Comparison with (\ref{eq:QHO}) shows that the density of states is itself a sum over squares of regularly spaced oscillator functions, albeit in energy space; their width $\Gamma$ and shifts $E_n$ are given by:
\begin{equation}
\label{eq:crossed34}
\Gamma = q\elf l = \elf\sqrt{\frac{q\hbar}\mgf} \;, \qquad E_n = E - (2n+1)\hbar\omega_L - mv_D^2/2 \;.
\end{equation}
Note that the centers of these oscillator states coincide with the Landau levels, apart from a constant shift due to the drift motion.  As $\elf\rightarrow 0$, the width $\Gamma$ tends towards zero, and the discrete energy levels familiar from a purely magnetic field emerge (\ref{eq:JB2D}).

\paragraph*{Propagator method}

To obtain the density of states (\ref{eq:crossed33}) using the propagator transform (\ref{eq:crossed31}), we must first extract the function $K_\perp(\mathbf o, t|\mathbf o,0)$ from (\ref{eq:crossed13}):
\begin{equation}
\label{eq:crossed35}
K_\perp(\mathbf o, t|\mathbf o,0) = \frac{m\omega_L}{2\pi\rmi\hbar\sin(\omega_L t)}
\exp \left\{ \frac{\rmi m v_D^2 t}{2\hbar} \bigl[ \omega_L t \cot(\omega_L t) -1 \bigr] \right\} \;.
\end{equation}
This expression formally resembles the propagator in a purely magnetic field (\ref{eq:KB}), and the generating function expansion (\ref{eq:LaguerreTheorem}) again yields the series expansion of $K_\perp(\mathbf o, t|\mathbf o,0)$.  As a function of the width parameter $\Gamma$ defined above (\ref{eq:crossed34}), it reads:
\begin{eqnarray}
\nonumber
K_\perp(\mathbf o, t|\mathbf o,0) & = & \frac{q\mgf}{2\pi\hbar} \,\rme^{-\Gamma^2t^2/(4\hbar^2)}
\sum_{n=0}^\infty \La_n \left( \frac{\Gamma^2 t^2}{2\hbar^2} \right) \\
& &\times \exp\left\{ -\frac{\rmi t}\hbar \left[ (2n+1)\hbar\omega_L + \frac{mv_D^2}2 \right]\right\} \;.
\label{eq:crossed36}
\end{eqnarray}
Indeed, the density of states in crossed fields (\ref{eq:crossed33}) follows after term-by-term integration of this sum in (\ref{eq:crossed31}), as can be shown using the Fourier transform \cite{Abramowitz1965a}:
\begin{equation}
\label{eq:crossed37}
\int_{-\infty}^\infty \rmd t\, \rme^{-(t-\rmi x)^2} \bigl[ \Her_n(t) \bigr]^2 =
2^n n! \sqrt\pi\, \La_n(2x^2) \;.
\end{equation}
For a more detailed discussion of this approach, see \cite{Kramer2003a}.  While the method appears unnecessarily complicated for the determination of $n_{\ELF\times\MGF}^{(2D)}(\mathbf o;E)$, the propagator formalism clearly offers an advantage when evaluating the Green function $G_{\ELF\times\MGF}^{(2D)}(\mathbf r_\perp,\mathbf o;E)$ for $\mathbf r_\perp \neq \mathbf o$, since a single numerical integration (\ref{eq:GEnergyIntegral}) will suffice to obtain the complete Green function.  We will show examples below.

\paragraph*{Canonical transformation method}

The result gained by the two previous methods is also consistent with the mapping of the original Hamiltonian in crossed fields in equation~(\ref{eq:HEB}) to the Hamiltonian of a shifted harmonic oscillator. Details of the corresponding canonical transformation and its unitary representation are discussed in \cite{Kramer2000a,Kramer2003b}.

\paragraph*{Properties of the density of states functional}

The functional form of the two-dimensional density of states (\ref{eq:crossed33}) in crossed fields is a major topic in Refs.~\cite{Kramer2003b,Kramer2003a}. Here we content ourselves with a short summary of the main features.
\begin{figure}[t]
\begin{center}
\includegraphics[width=0.49\textwidth]{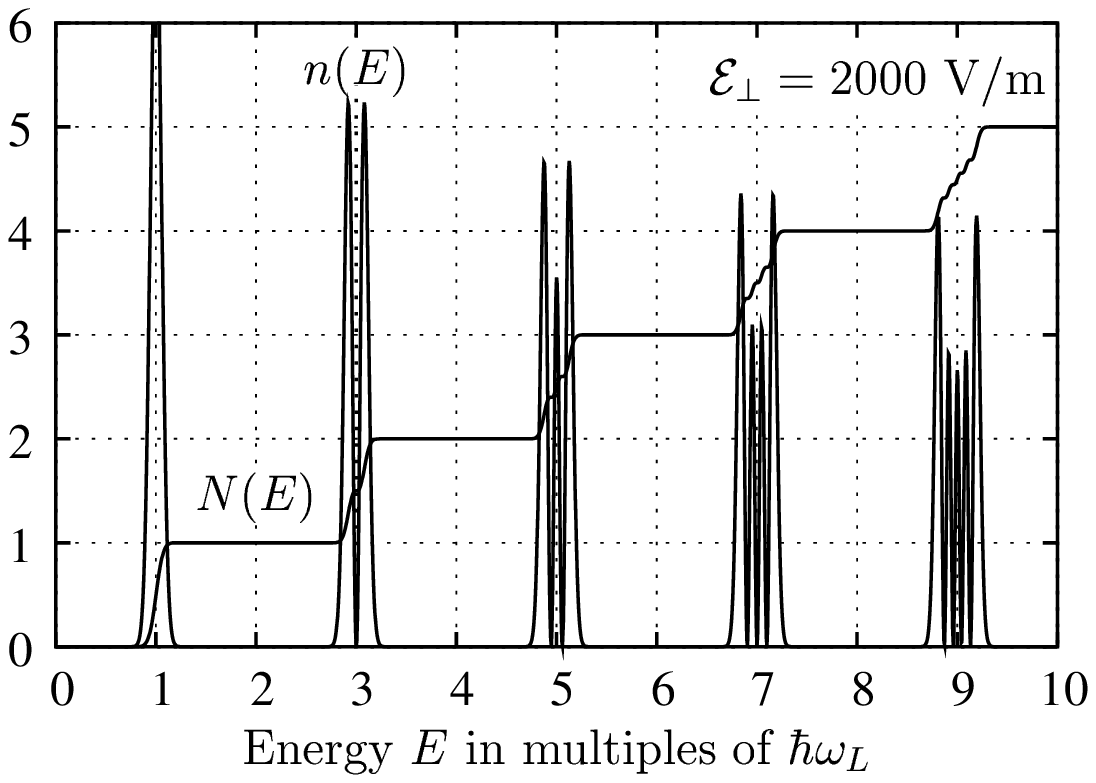}
\includegraphics[width=0.49\textwidth]{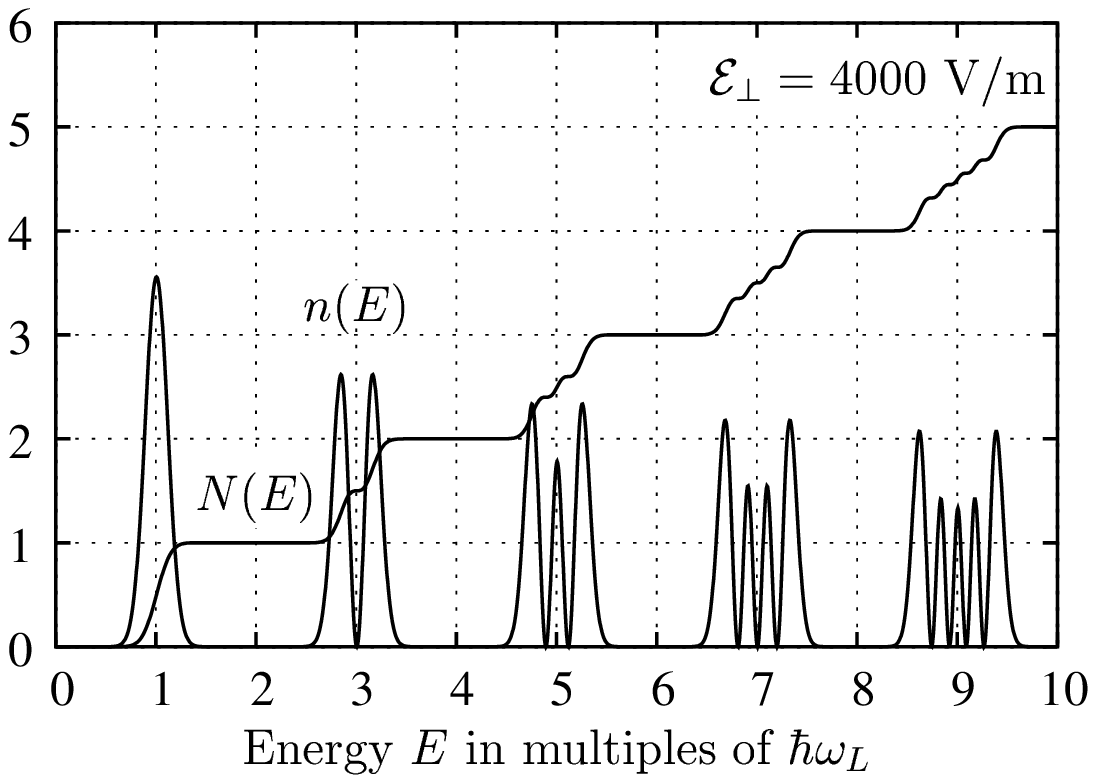}\\
\includegraphics[width=0.49\textwidth]{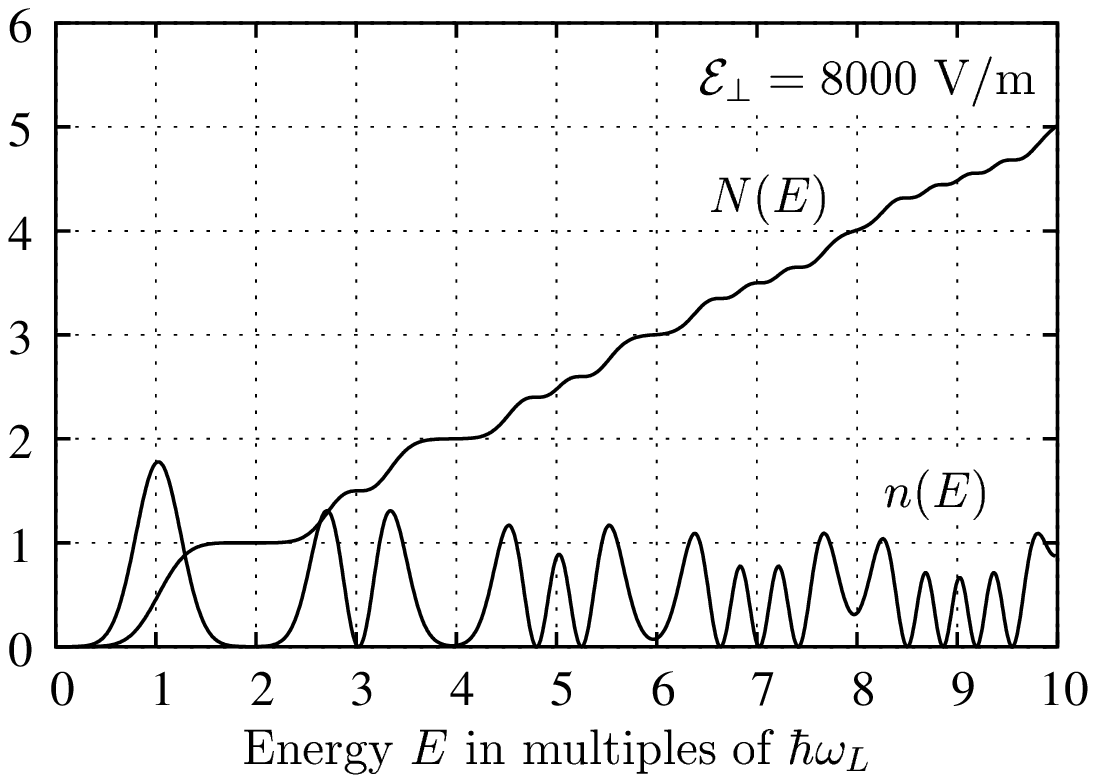}
\includegraphics[width=0.49\textwidth]{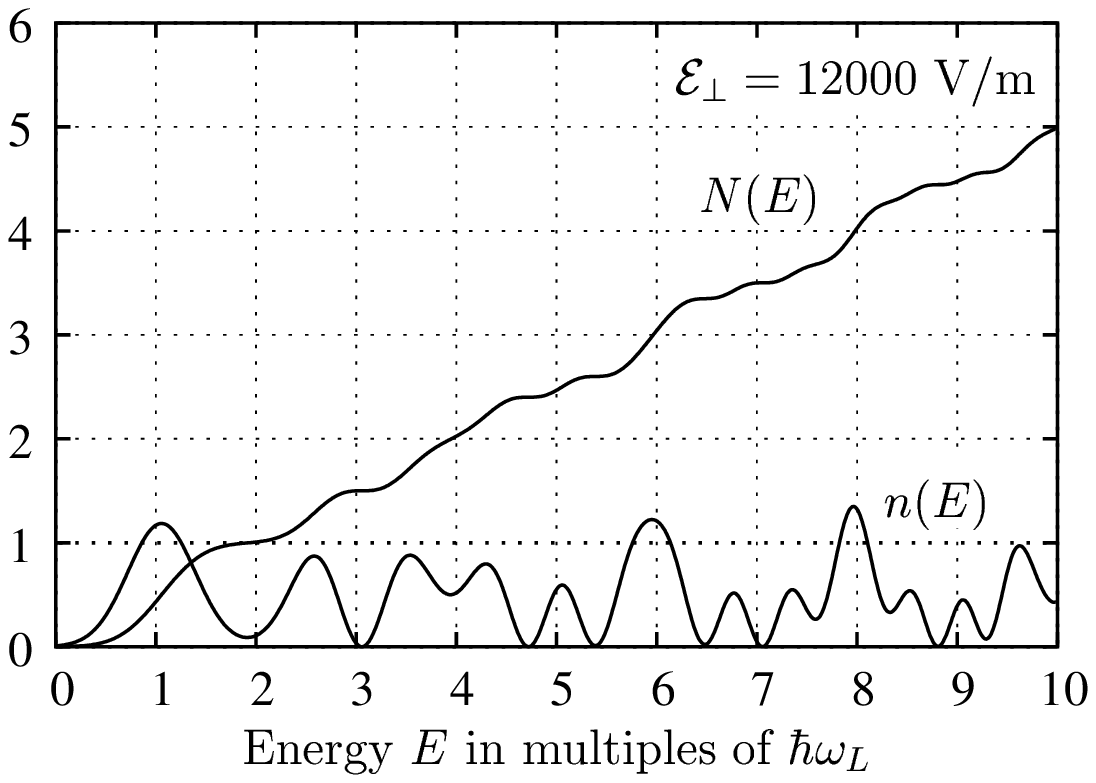}
\end{center}
\caption[2D electronic density of states in crossed fields]{Two-dimensional local density of states (LDOS) $n(\mbfo;E)$ (in units of $m /(\pi\hbar^2)$) and integrated LDOS $N(\mbfo;E)$ (in units of $q\mgf /(2\pi\hbar)$) at four different electric fields $\Elf=2000, 4000, 8000,12000$~V/m and for a magnetic field $\mgf =5$~T as a function of the scaled energy $E/(\hbar\omega_L)$ according to equation~(\ref{eq:crossed33}). Near the $n$th Landau level at $E=(2n+1)\hbar\omega_L$, the DOS renders the probability distribution of a one-dimensional harmonic oscillator in the $n$th eigenstate.}\label{fig:DOSIDOS}
\end{figure}
As noted above, the density of states consists of a sum over equally weighted harmonic oscillator eigenstates that appear not in configuration space but as functions of the energy $E$.  As the eigenfunctions $u_n(\xi)$ (\ref{eq:QHO}) form an orthonormal set, the total contribution of each sum term, i.~e., each Landau level is given by:
\begin{equation}
\label{eq:EBQ}
\int_{-\infty}^\infty\rmd E\; n_{n,\ELF\times\MGF}^{(2D)}(E) = \frac{e\mgf}{2\pi\hbar} \;.
\end{equation}
This result is in accordance with the quantization of the Landau levels in a purely magnetic field (\ref{eq:JB2D}). For each Landau level $n$, the density of states has a Gaussian envelope with width $\Gamma$ that is split into $n+1$ intervals by the $n$ simple zeroes $\xi_{n,j}$ ($j=1,\ldots,n$) of the polynomial $\Her_n(\xi)$.  In Fig.~\ref{fig:DOSIDOS}, we plot the resulting density of states for various electric field strengths $\elf$.  For small $\elf$, the overlap between adjacent Landau levels is negligible, as the DOS drops off exponentially between them.  With increasing electric field, the Landau levels broaden and finally coalesce.  We infer from equation~(\ref{eq:QHO}) that the classical turning point of harmonic motion, $\xi^{\rm tp}_n=\sqrt{2n+1}$, provides a practical measure for the width of the partial density of states $n_{n,\ELF\times\MGF}^{(2D)}(E)$.  The populated region in energy between adjacent Landau levels $n-1$, $n$ is then approximately given by the ratio:
\begin{equation}
\label{eq:LLoverlap}
\frac{{\rm combined~half~widths~of~levels}\;\Gamma(\xi^{\rm tp}_{n-1}+\xi^{\rm tp}_n)}{{\rm level~spacing}\;2\hbar\omega_L}
\sim 2\sqrt{2n} \, \frac{m\elf}{\sqrt{q\hbar\mgf^3}} \;.
\end{equation}
The overall extension of the modulated Landau levels increases with $n^{1/2}$ for fixed fields.  Note that all features of $n_{n,\ELF\times\MGF}^{(2D)}(E)$, including the nodes, scale linearly in width with the electric field $\elf$.  While for small ratios in (\ref{eq:LLoverlap}) individual levels remain well separated, with increasing overlap the density of states becomes a smooth function of the energy $E$. This transition is clearly visible in Figure~\ref{fig:DOSIDOS}.
\begin{figure}[t]
\centerline{
\includegraphics[height=1.67in]{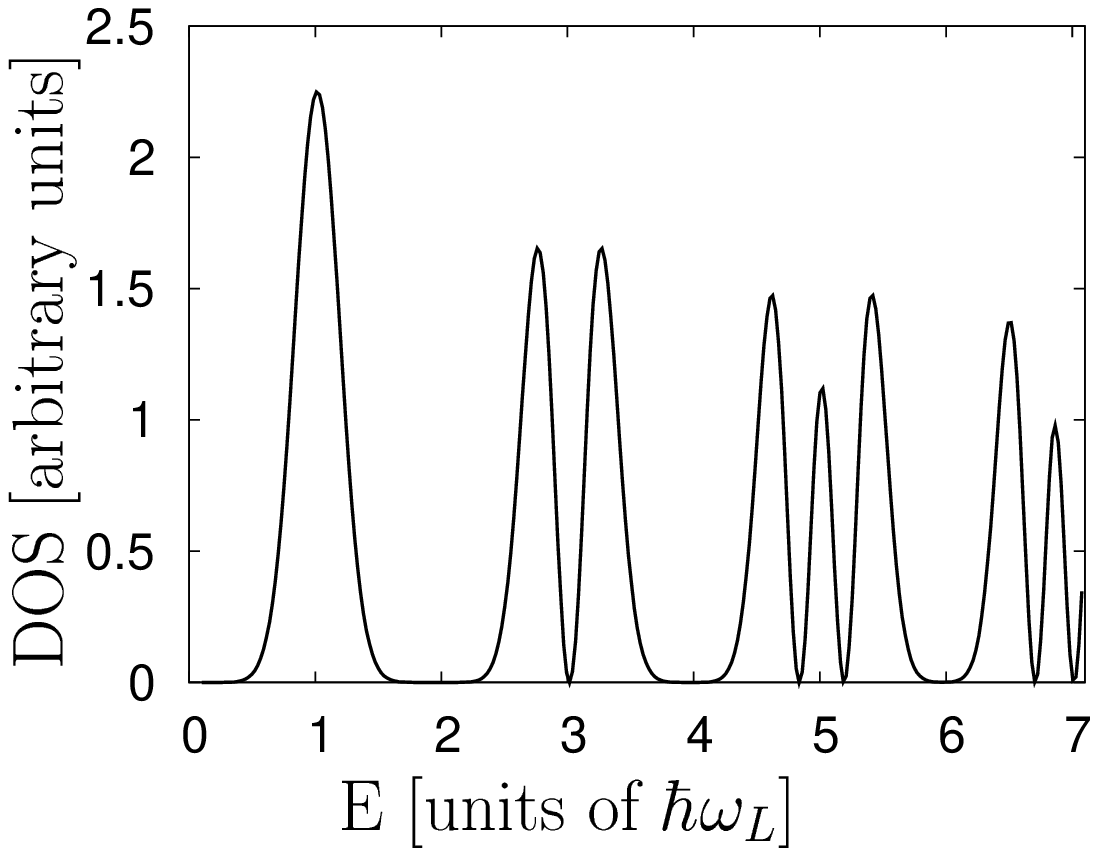}\quad
\includegraphics[height=1.67in]{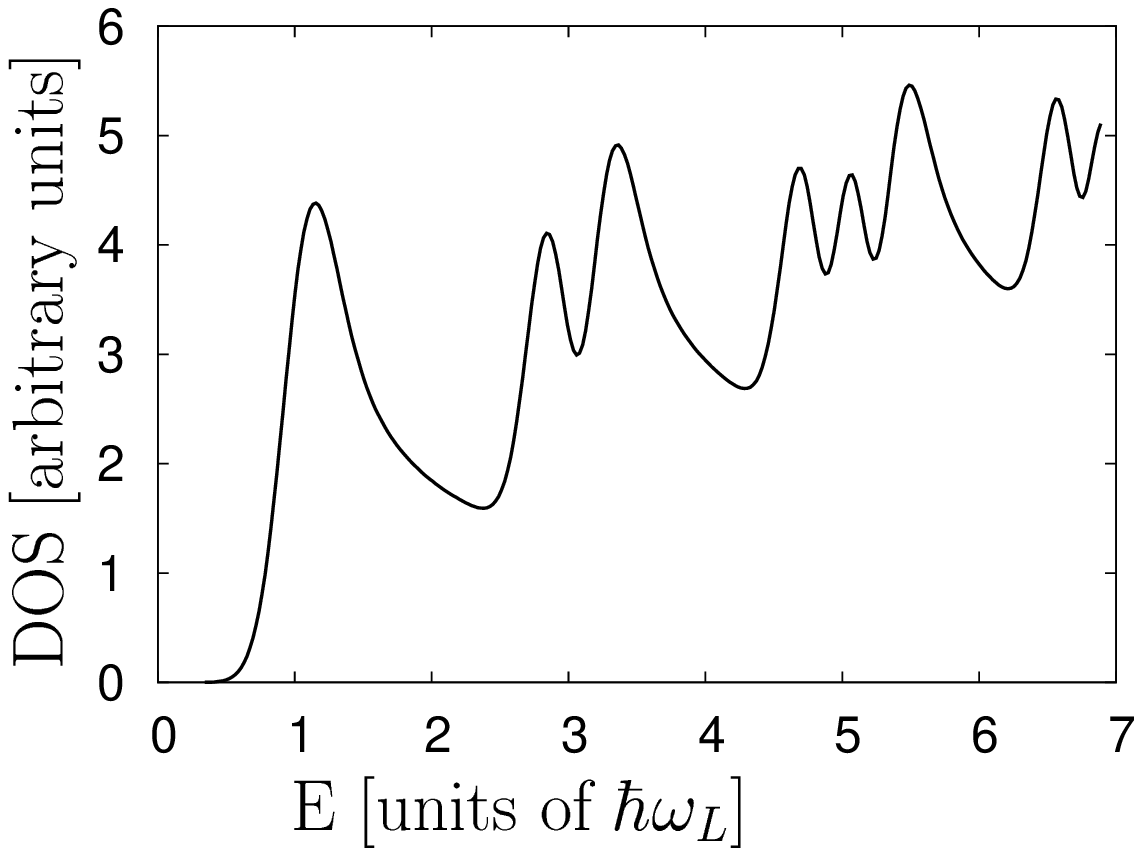}
}
\caption[Density of states in an electron gas with magnetic and electric field]
{Electronic density of states in two (left panel) and three dimensions (right panel) in perpendicular electric and magnetic fields ($\mgf=0.5$~T, $\elf=200$~V/m). }\label{fig:DOSEB}
\end{figure}

\subsubsection{Extension to three dimensions}

In three dimensions, besides the Green function $G_{\ELF,\MGF}(\mbfr,\mbfo;E)$ even the density of states functional $n_{\ELF,\MGF}^{(3D)}(\mathbf o;E)$ defies evaluation in closed form.  However, simple integral representations are available.  Starting from the identity (\ref{eq:crossed31}), we may employ the composition property (\ref{eq:crossed11}) in order to obtain an integral representation:
\begin{equation}
\label{eq:crossed38}
n_{\ELF,\MGF}^{(3D)}(\mathbf o;E) = \frac1{2\pi\hbar}\, \int_{-\infty}^\infty \rmd T\, \rme^{\rmi ET/\hbar}\, K_\perp(\mathbf o, T|\mathbf o,0) K_\parallel(0, T|0,0) \;.
\end{equation}
Alternatively, we may formally perform the integration in (\ref{eq:crossed38}), which leads to a convolution integral of the individual transforms, similar to (\ref{eq:ComplexConvolution}).  This approach yields a simple composition theorem for the density of states:
\begin{equation}
\label{eq:crossed39}
n_{\ELF,\MGF}^{(3D)}(\mathbf o;E) = \int_{-\infty}^\infty \rmd E'\,
n_{\ELF\times\MGF}^{(2D)}(\mathbf o;E') n_{\Elf_\parallel}^{(1D)}(0;E-E') \;.
\end{equation}
While the former expression is better suited for numerical calculations, (\ref{eq:crossed39}) yields more physical insight:  Inserting the series expansion (\ref{eq:crossed33}) into (\ref{eq:crossed39}),we infer that the three-dimensional density of states can again be interpreted as a sum over individual Landau levels $n$, where their actual contribution $n_{n,\ELF,\MGF}^{(3D)}(\mathbf o;E)$ follows from convolution of the oscillator function $\bigl[ u_n(E_n/\Gamma) \bigr]^2$ (\ref{eq:QHO}) with the one-dimensional density of states $n_{\ELF_\parallel}^{(1D)}(0;E-E')$.  (In the case of perpendicular fields ($\Elf_\parallel = 0$), these integrals can be expanded into series of parabolic cylinder functions \cite{Blumberg1979a}, but we will not elaborate this point further.)  In Figure~\ref{fig:DOSEB} we compare the analytic two-dimensional solution (\ref{eq:crossed33}) and the corresponding three-dimensional density of states (\ref{eq:crossed38}) in orthogonal fields.  The close relation between both functionals is clearly displayed, as well as the separation of $n_{\ELF,\MGF}^{(3D)}(\mathbf o;E)$ into individual Landau levels.

\subsection{Spin}
\label{subsec:spin}

A slight complication occurs if the motion of charges with spin, like electrons, is considered, since the spin interacts with the magnetic field $\MGF$.  However, for uniform magnetic field, this interaction merely causes a constant effective energy shift $\Delta E = \pm \frac{1}{2}\, g\hbar\omega_L$ if we select the magnetic field direction as axis of quantization.  Thus, the Green functions for each spin component follow from its scalar counterpart by adjusting their energy, $G_{\uparrow,\downarrow}(\mathbf r,\mathbf r';E) = G(\mathbf r,\mathbf r';E \pm \Delta E)$.  Similarly, the spin dependent densities of states become
\begin{equation}
\label{eq:spin}
n_{\uparrow,\downarrow}(E) =n\left(E \pm \frac{1}{2}g\hbar\omega_L\right) \;,
\end{equation}
and the total density of states including spin can be mapped back to the scalar quantity: $n_{\uparrow\downarrow}(E) = n_\uparrow(E) + n_\downarrow(E)$.  Hence we defer the inclusion of spin for the moment.

\section{Application: Photodetachment}
\label{sec:photo}

In a photodetachment experiment, electrons are detached from negatively charged ions $A^-$ due to the interaction with a laser field:
\begin{equation}
\label{eq:photo0.1}
A^- + h\nu \rightarrow A + e^- \;.
\end{equation}
The detached electron is emitted with a definite energy $E$ given by the difference between its binding energy (or affinity) and the photon energy. This process allows a description in terms of quantum sources.
In near-threshold detachment ($E\rightarrow0$), it is reasonable to model the ion as a point source because its size is small compared to the de~Broglie wavelength of the emitted electron. (Here, we consider only the generation of $s$--waves.  For the general case of multipole emission, see Ref.~\cite{Bracher2003a}.)
The photodetachment current in external fields is then linked to the relevant energy-dependent Green function: For a laser beam illuminating the ions for the duration $T$, their survival probability is given by
\begin{equation}
R(E)\propto\exp[-J(E) T],
\end{equation}
where $J(E)$ denotes the total current defined in equation~(\ref{eq:JDelta}).  In practice, an external electric field can provide a virtual double-slit environment that allows to probe the energy of the emitted electron (and thereby the electron affinity of the ion) with extreme accuracy \cite{Blondel1996a,Blondel1999a,Bracher2003a}.

The combination of electric and magnetic fields imprints a non-trivial structure on the detachment rate and allows to identify features of the underlying energy-dependent Green function.  Unfortunately a direct experimental observation of these features is obscured by several effects. Typically, the negative ions are confined in an ion trap, where they still have a large kinetic energy. In a thermal ion cloud the momentum distribution $P(\mathbf{p})$ is given by Maxwell's expression:
\begin{equation}
\label{eq:th5}
P(\mathbf p)=
\frac{1}{{(2\pi m k_B T)}^{3/2}}\exp\left(-p^2/(2 m k_B T)\right).
\end{equation}
In an external magnetic field, the charges will experience an electric field in their rest frame due to the transformation of the fields \cite{Jackson1975a} that accounts for the Lorentz force,
\begin{equation}
\label{eq:Fl}
\elf(\mathbf{p})=\frac 1m\, \mathbf{p} \times \MGF.
\end{equation}
This electric field is exactly perpendicular to the momentum and the external magnetic field. A stationary source will only emerge if we consider the photon-electron interaction in the rest frame of the ion.  Hence, we employ the three-dimensional Green function for crossed electromagnetic fields to describe the photodetachment of moving ions in a purely magnetic field.
\begin{figure}[t]
\begin{center}
\includegraphics[width=0.7\textwidth]{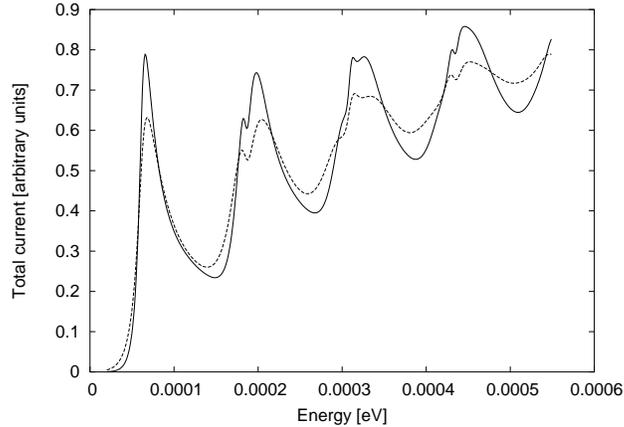}
\end{center}
\caption[Photodetachment in crossed fields at different temperatures.]{Thermally averaged curves for the total photocurrent as a function of the electron energy. Magnetic field: $\mgf=1.07$~T, ion mass: $m=32$~u. Solid line: $T=400$~K, dashed line $T=950$~K (see also \cite{Blumberg1979a}, Figure~2). The substructure and broadening of the Landau levels due to the perpendicular electric field is visible. However, the features are washed out (compared to Figure~\ref{fig:DOSEB}) due to the averaging over a wide range of electric field values.
}\label{fig:Baverage}
\end{figure}

The averaging effect of varying electric fields due to the thermal motion is displayed in Figure~\ref{fig:Baverage}. A comparison with the plot for a single value of the electric field (right panel in Figure~\ref{fig:DOSEB}) shows that the substructure of the Landau levels changes. Some features are still visible, like the division of the first level due to the zero of the first Hermite polynomial.
\begin{figure}[t]
\begin{center}
\includegraphics[width=0.7\textwidth]{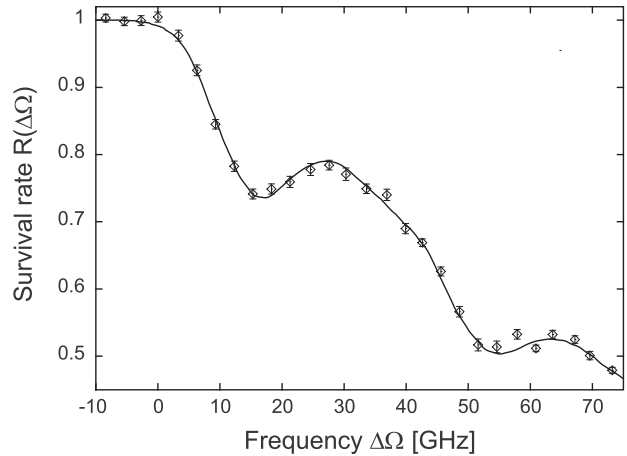}
\end{center}
\caption{Ratio of surviving ions $R(\Delta\Omega)$ in photodetachment of S$^-$ in an external magnetic field as a function of the laser detuning $\Delta\Omega$. The solid line is the theoretical prediction, the circles represent experimental data \cite{Yukich2003a}. Parameters used:
$\mgf=1$~T, ion mass $m=32$~amu, $T=2100$~K.}
\label{fig:Brate}
\end{figure}
Another complication stems from the Zeeman splitting of the ionic energy levels in an external magnetic field. Experimentally, usually a superposition of many allowed transitions is observed.  A recent experiment is compared to the theory sketched here in Ref.~\cite{Yukich2003a}. As shown in Figure~\ref{fig:Brate} the agreement is excellent and underlines the validity of the quantum source approach. An alternative theoretical description, together with earlier experimental data is put forward in~\cite{Blumberg1979a,Blumberg1978a}.

\section{Application: Quantum Hall effect}
\label{sec:QHE}

\begin{figure}[t]
\centerline{\includegraphics[width=0.5\textwidth]{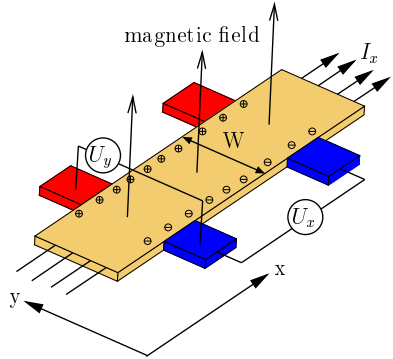}}
\caption{Schematic picture of a Hall bar. A constant current $I_x$ is flowing along the $x$-axis. Perpendicular to the current and an external magnetic field, the Hall field is established along the $y$-axis to counterbalance the deflection of the electrons. Experiments record the Hall potential $U_y$ and the longitudinal potential $U_x$.
}\label{fig:Hallbar}
\end{figure}
Another application of the Green function in crossed fields is the quantum Hall effect in a two-dimensional electron gas (2DEG). In Figure~\ref{fig:Hallbar} we show the basic geometry of the sample. In the system, a constant current is sent along the $x$-axis of the sample. Perpendicular to the surface of the electron gas a strong magnetic field is applied. In a classical picture, initially electrons entering the sample are deflected to one edge, and a potential across the sample builds up until the Lorentz force is compensated by the induced electric field. The electrons then drift in the crossed fields with the constant velocity $\mathbf v_D$ (\ref{eq:crossed14}) perpendicular to both fields. (Note that this mechanism leads to a loss-free stationary current in the presence of an electric field, unlike conventional transport theory, where the current is limited by inelastic scattering instead.)  The linear relation between electric field and current density in two dimensions is expressed by the conductivity tensor $\mbfsigma$:
\begin{equation}\label{eq:conductivitymatrix}
\left(
\begin{array}{c}
j_x\\
j_y
\end{array}
\right)
=
\left(
\begin{array}{cc}
\sigma_{xx} & \sigma_{xy} \\
\sigma_{yx} & \sigma_{yy}
\end{array}
\right)
\left(
\begin{array}{c}
\Elf_x\\
\Elf_y
\end{array}
\right).
\end{equation}
Its inverse, the resistivity tensor $\mbfrho$, is related to the conductivity via
\begin{equation}
\label{eq:qh00}
\rho_{xy} = \frac{\sigma_{yx}}{\sigma_{xx}^2+\sigma_{xy}^2},\quad\quad
\rho_{xx} = \frac{\sigma_{xx}}{\sigma_{xx}^2+\sigma_{xy}^2}.
\end{equation}
We remark that for $\sigma_{xy} \neq 0$, vanishing resistivity $\rho_{xx}$ implies vanishing conductivity $\sigma_{xx}$.  In the setup shown in Figure~\ref{fig:Hallbar}, for stationary current density $j_x$ and  transverse electric field $\Elf_y$, the current $j_y$ has to be zero. We will now explore different models for the conductivity in the two-dimensional Hall effect.

\subsection{Drift transport of electrons}

\begin{figure}[t]
\centerline{
\includegraphics[width=0.5\textwidth]{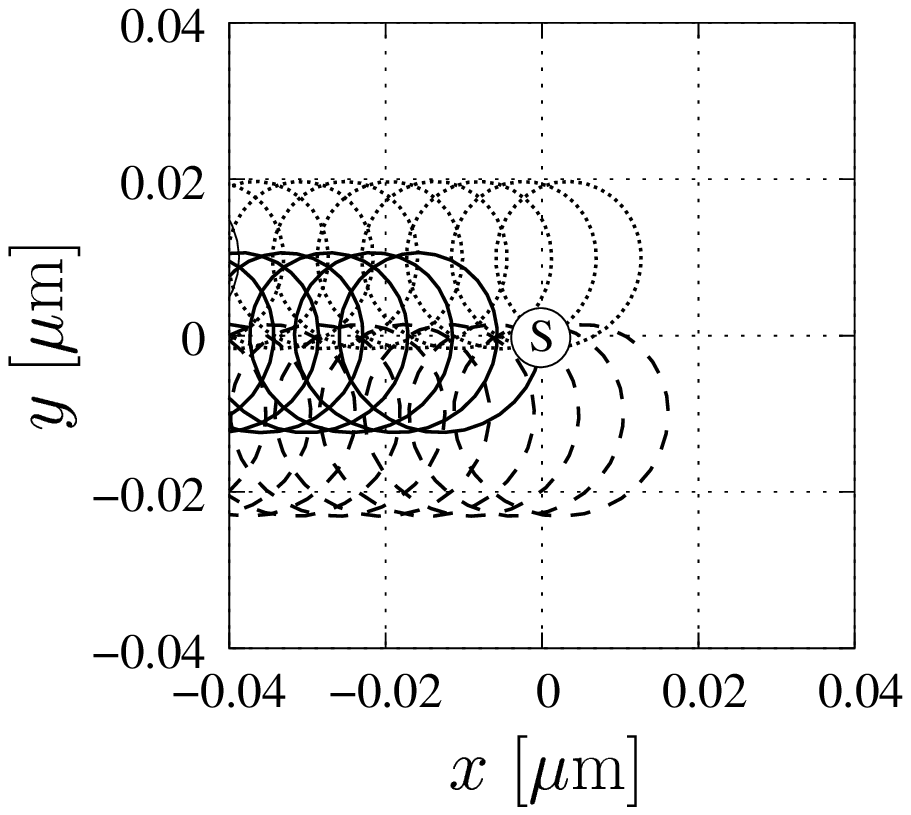}
\includegraphics[width=0.5\textwidth]{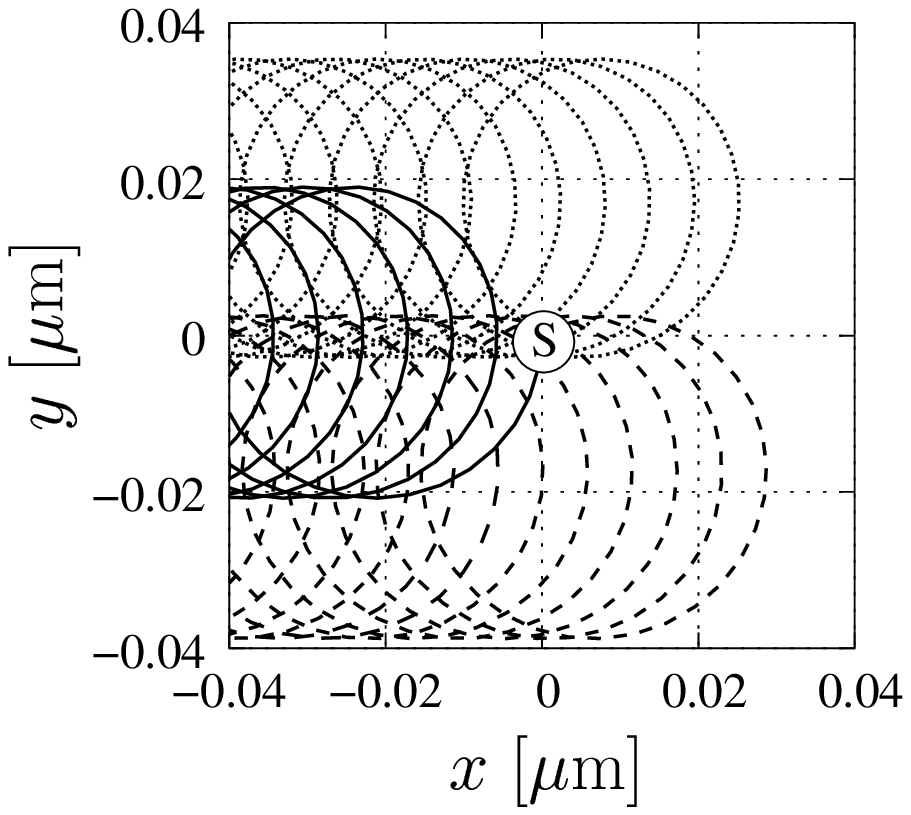}
}
\caption{Classical trajectories from a point source $\bigo{S}$ located at $x=y=0$. The magnetic field $\mgf=5$~T is oriented perpendicular to the plane, the electric field $\Elf=4000$~V/m along $\mbfe_y$. Left panel: Energy $E=\hbar\omega_L$, right panel: $E=3\hbar\omega_L$. While the radius of cyclotron motion varies, the average drift velocity $v_D=\Elf/\mgf$ is the same in both panels.}
\label{fig:classicaltraj}
\end{figure}

\subsubsection{Classical transport}

In this section we will review the classical transport of electrons in a Hall sample. We will take a modest level of scattering of the conduction electrons into account. In a simple Drude-like model the dynamics of the electrons is governed by the Lorentz force, amended for a term that incorporates friction via a relaxation time $\tau$:
\begin{equation}
m\frac{\rmd \mbfv}{\rmd t}=e\ELF+e\mbfv\times\MGF-\frac{m}{\tau}\mbfv.
\end{equation}
Under stationary conditions $\mbfv$ is constant and together with the current density $\mbfj=N e \mbfv$ the components of the conductivity tensor in equation~(\ref{eq:conductivitymatrix}) become
\begin{equation}
\label{eq:sigmaclassical}
\mbfsigma=
\frac{e^2 N\tau}{m}
\frac{1}{1+\omega_C^2\tau^2}
\left(
\begin{array}{cc}
1&-\omega_C\tau \\ \omega_C\tau&1
\end{array}\right),
\end{equation}
where $\omega_C=2\omega_L=e\mgf /m$. Inverting this matrix we extract the resistivity components
\begin{equation}
\label{eq:rho_cl}
\rho_{xy}=\frac{\mgf}{N e},\qquad
\rho_{xx}=\frac{m}{N e^2\tau}.
\end{equation}
In order to connect this picture to the classical drift of electrons in crossed fields we use the relation
\begin{equation}
\Elf_y=\rho_{xy}j_x=\frac{\mgf}{N e} j_x,\qquad j_y=0.
\end{equation}
Solving for $v_x^{cl}=j_x/(eN)$ yields $v_x^{cl} = v_D = \Elf_y/\mgf$:  The current density along the $x$-direction is given by the electron density $N$, multiplied by the drift velocity $v_D$. Some classical electron trajectories are shown in Figure~\ref{fig:classicaltraj}. As we will see in the next section, the quantum mechanical picture radically diverts from these results.
\begin{figure}[t]
\centerline{\includegraphics[width=1.0\textwidth]{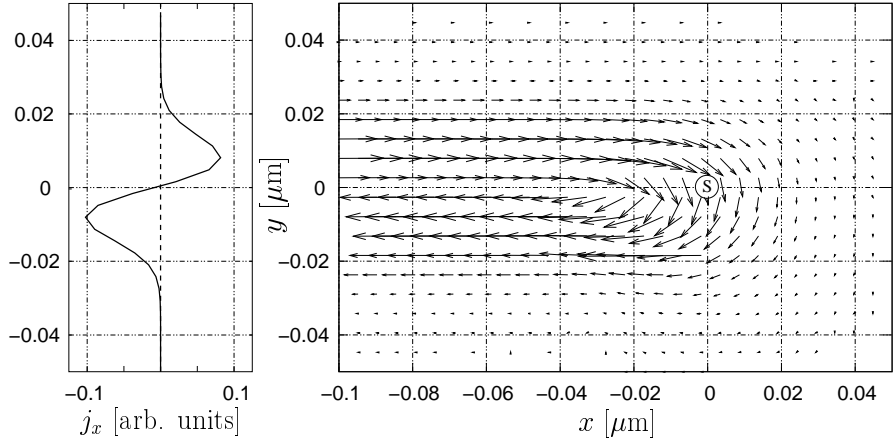}}
\caption{Current density from a point source $\bigo{S}$ located at $x=y=0$ in  crossed fields $\mgf=5$~T, $\Elf_y=4000$~V/m (cf.\ Figure~\ref{fig:classicaltraj}). The electron energy is $E=\hbar\omega_L$. Left panel: The component $j_x(x,y)$ at $x=-0.1\;\mu$m. Right panel: Spatial current flow. The arrows indicate the direction of the current and their length is proportional to $|j|^{1/4}$.}
\label{fig:qmll1}
\end{figure}

If we assume a field independent carrier density $N$, equation~(\ref{eq:rho_cl}) leads to the classical Hall effect: The Hall resistivity $\rho_{xy}$ is linearly dependent on the magnetic field $\mgf$, and the constant of proportionality renders the number of carriers that participate in the transport.

\subsubsection{Quantum mechanical drift}\label{sec:qmdrift}

\begin{figure}[t]
\centerline{\includegraphics[width=0.8\textwidth]{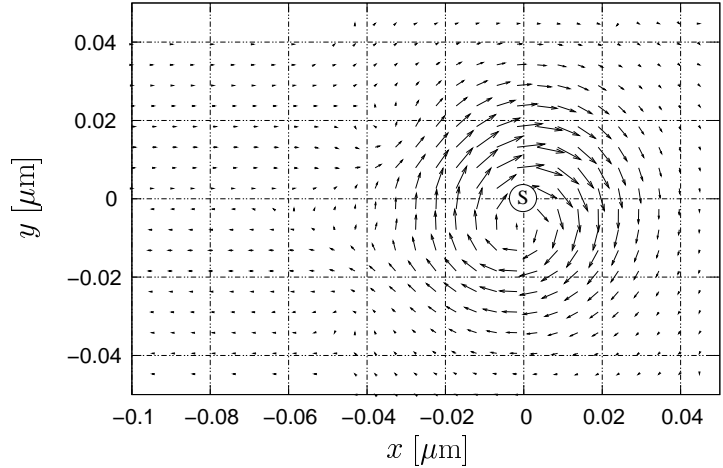}}
\caption{Current density from a point source $\bigo{S}$ located at $x=y=0$ in crossed fields $\mgf=5$~T, $\Elf_y=4000$~V/m (cf.\ Figure~\ref{fig:classicaltraj}). The electron energy is $E=3\hbar\omega_L$. The arrows indicate the direction of the spatial current flow and their length is proportional to $|j|^{1/4}$.}
\label{fig:qmll3}
\end{figure}

What do the quantum current and transport look like in our quantum source model of electronic matter waves? Equations~(\ref{eq:currdens}) and~(\ref{eq:JDelta}) yield quantum mechanical expressions for the currents originating from a point source. In Figures~\ref{fig:qmll1} and \ref{fig:qmll3} we plot the spatial current distribution generated in a magnetic field of $\mgf=5$~T and an electric field of $\Elf_y=4000$~V/m. In the first plot we chose $E=\hbar\omega_L$ which corresponds to the first maximum in the density of states (\ref{eq:crossed33}) (see Figure~\ref{fig:DOSEB}), whereas in the second plot the energy $E=3\hbar\omega_L$ is close to a minimum in the total current.  Some corresponding classical trajectories are shown in Figure~\ref{fig:classicaltraj}. The quantum mechanical current distribution shows some intriguing features: In the vicinity of the source (located at the origin), a complicated flow pattern emerges. At some distance from the source, the current follows the classical drift direction, but is split into two stripes with anti-parallel current vectors. We will discuss the implications of these oppositely flowing currents again in Section~\ref{sec:FlowPattern}.

\begin{figure}[t]
\centerline{\includegraphics[width=0.8\textwidth]{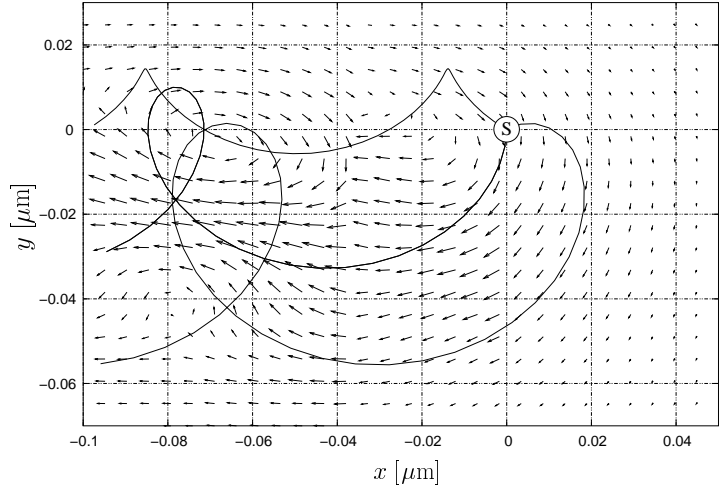}}
\caption{Current density from a point source $\bigo{S}$ located at $x=y=0$ in perpendicular fields $\mgf=2$~T, $\Elf_y = 8000$~V/m. The electron energy is $E=\hbar\omega_L$. The arrows indicate the direction of the current flow and their length is proportional to $|j|^{1/4}$. Three classical trajectories of electrons with the same energy are also shown.}
\label{fig:flowlow}
\end{figure}

The current distribution for stronger Hall fields is depicted in Figure~\ref{fig:flowlow}. Here the drift velocity is $v_D = 4000$~m/s. Three classical trajectories are included in the plot. The nearly circular orbits of Figure~\ref{fig:classicaltraj} are distorted to trochoidal shapes \cite{Peters1993a}, and also the quantum mechanical current profile changes considerably.

\paragraph*{Total current vs.\ current density}

The equation of continuity is valid for any surface enclosing the point source, and therefore the spatial current density integrated over such a closed surface must yield the total current, which by (\ref{eq:JDOS}) is proportional to the density of states (\ref{eq:crossed33}). Since the total current is available in analytic form, we may use this relation to cross-check our numerical evaluation of the spatial current density.  As a function of $E$, the functional (\ref{eq:crossed33}) repeatedly virtually drops to zero.  This almost vanishing total current does not imply vanishing current density, however, as Figure~\ref{fig:qmll3} demonstrates: Circular flow patterns lead to a very small net-flow from the source region. Further away the contributions of oppositely directed flows cancel each other almost perfectly.

\paragraph*{Recovering the drift velocity}

In order to define an average velocity along the drift direction we proceed as follows: For a slice along the $y$-axis at some fixed distance $x$ from the source we calculate the integrated density $\varrho(x;E)$ using
\begin{equation}\label{eq:varrho}
\varrho(x;E) = \int_{-\infty}^{+\infty}\rmd y\;{|G(\mbfr=(x,y),\mbfo;E)|}^2
\end{equation}
and then define the ratio
\begin{equation}
v_x^{\rm av}(x;E) = J(\mathbf o;E)/\varrho(x;E)
\end{equation}
as the average velocity. This procedure yields values very close to the classical drift velocity $v_D = 800$~m/s in both cases illustrated. However, $J(\mathbf o;E)$ and $\varrho(x;E)$ are drastically different for the two energies chosen in Figures~\ref{fig:qmll1} and \ref{fig:qmll3}. (We note that the local velocity field $\mathbf v(\mathbf r) = \mathbf j(\mathbf r) / |G(\mbfr=(x,y),\mbfo;E)|^2$ greatly varies with $\mathbf r$.  Only the integrated quantity reproduces the drift.)  Furthermore, it is important to realize that the intensity of the current $J(\mathbf o;E)$ is exponentially suppressed at certain energies as shown in Figure~\ref{fig:DOSEB}. This is in sharp contrast to the classical picture, where a constant drift transport occurs for all energy values of the injected electrons.

\subsection{Fermionic matter waves}
\label{sec:fermionicmw}

The density of states is a single-particle quantity. In a solid, many electrons take part in the conduction process. For a non-interacting system, the available single-particle energy levels are occupied according to Fermi-Dirac statistics. Taking spin into account, two electrons may share each state. For a system that exhibits a point spectrum of the energy levels (e.g.\ atoms, or a purely magnetic field in two dimensions (\ref{eq:JB2D})), the resulting electronic configuration is similar to the shell structure of atoms. For a continuous spectrum, the Fermi energy controls the integrated carrier density of the system, which at temperature $T\rightarrow 0$ is given by the integrated density of states $N(\mbfo;E_F)$:
\begin{equation}
\label{eq:IDOS}
N(\mbfo;E_F,\Elf_y,\mgf)=\int_{-\infty}^{E_F}\rmd E\;n(\mbfo;E,\Elf_y,\mgf).
\end{equation}
In crossed electric and magnetic fields, (\ref{eq:IDOS}) is available in closed form; see Ref.~\cite{Kramer2003b}, Appendix~C. Clearly, $N(\mbfo;E_F)$ depends on the external magnetic and electric fields.

Here, we should point out that in multi-electron systems Coulomb-type interactions will occur. In the following we will ignore these interactions. The Pauli principle (which requires antisymmetric wave-functions for fermions) ensures that two electrons sharing the same spin will not be at the same position. Moreover scattering events that redistribute the electrons to different energy states take place only if the process involves filled initial and empty final states. At $T\rightarrow 0$, these states are available only close to the Fermi level of the system.

\subsection{Fermi energy in open and closed system}

\begin{figure}[t]
\centerline{\includegraphics[width=0.9\textwidth]{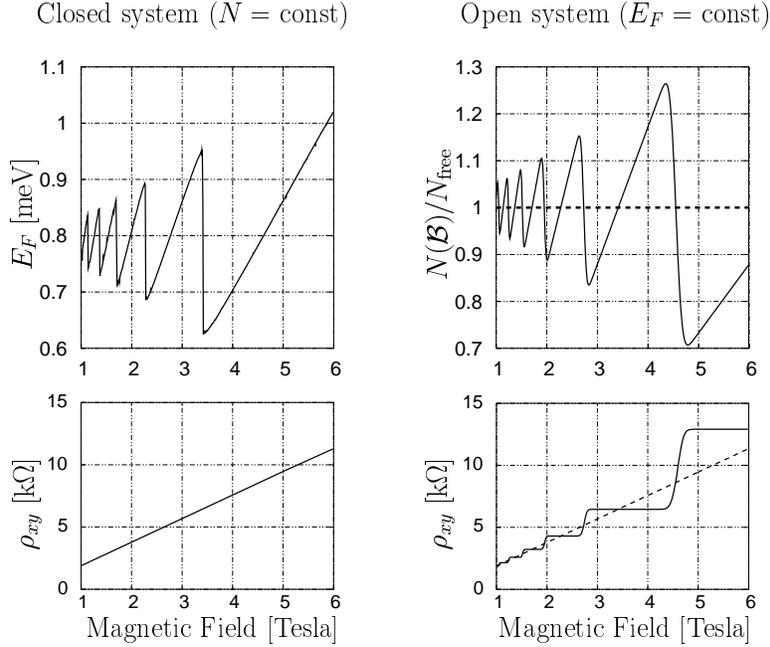}}
\caption{Top left: Fluctuation of the Fermi energy as a function of the magnetic field for fixed carrier concentration. Top right: Fluctuation of the current carrier concentration as a function of the magnetic field for fixed Fermi energy. Lower panels: Corresponding resistivity plots (the dashed line denotes the average number of carriers).}
\label{fig:fluc}
\end{figure}

\paragraph*{Closed system}

If we treat the two-dimensional electron gas as a system that is closed and decoupled from reservoirs, the number of electrons is a fixed quantity. The Fermi energy is determined from the relation
\begin{equation}
N=\mathrm{const}=N(\mbfo;E_F,\Elf_y,\mgf)=\int_{-\infty}^{E_F}\rmd E\;n(\mbfo;E,\Elf_y,\mgf).
\end{equation}
Changing the external fields (and therefore the density of states $n(\mbfo;E)$) leads to jumps in the Fermi energy as depicted in Figure~\ref{fig:fluc} (top left panel): The upper limit of the integral has to be adjusted in order to keep the number of carriers $N$ constant. Another implication of constant carrier density is a linear relationship between $\rho_{xy}$ and $\mgf$ (\ref{eq:rho_cl}). Thus different mechanisms have to be invoked in order to explain the existence of finite Hall ``plateaus'' of constant conductivity in the closed system. Proposals include the formation of one-dimensional conduction channels along the edges of the sample, and disorder. In a one-dimensional device Landauer quantization gives rise to discrete values of the conductivity. Disorder is supposed to lead to localized states populated by electrons which do not participate in the transport but nevertheless allow to adjust the Fermi energy smoothly \cite{Hajdu1994a}.

\paragraph*{Open system}

In the following we consider the implications of an open Hall system, where electrons can enter and leave the system through the contacts. In this picture the Fermi energy is fixed, while the number of particles fluctuates around the average free-particle value observed for $\mgf=0$. In Figure~\ref{fig:fluc} (upper right panel) we plot the oscillations of
\begin{equation}
N(\mbfo;E_F,\Elf,\mgf)/N_{{\rm free}}^{(2D)}(E_F)
\end{equation}
as a function of the magnetic field $\mgf$.  For comparison, we also show the resulting Hall resistivity $\rho_{xy}=\mgf/(N e)$ in Figure~\ref{fig:fluc} (lower right panel) together with its classical counterpart obtained by using $N = N_{{\rm free}}^{(2D)}(E_F)$ (\ref{eq:JFree}).  The following section is devoted to a detailed discussion of this curve. Here, we merely note that the difference in carrier density vanishes at the intersection points of both resistivity curves.  Otherwise, excess charges will be present whose electrostatic interaction will lead to potentials that subsequently alter the Fermi energy of the system. In the present discussion we will neglect this feedback mechanism.

\paragraph*{Experimental evidence for fluctuations}

Experiments show two types of fluctuations in quantum Hall systems as a function of the external magnetic field: 

Density fluctuations are directly observed in \cite{Raymond1999a} and fit well into the picture of an open system. These results contradict the basic theoretical assumptions for quantum Hall systems in Ref.~\cite{Hajdu1994a,Yoshioka2002a}, where $N={\rm const}$ is used to determine the currents. 

According to another experiment \cite{Wei1997a,Weitz2000a} the electrostatic potential measured atop the two-dimensional quantum Hall system fluctuates as a function of the magnetic field. These changes are interpreted by the authors as variations of the (local) chemical potential. However, according to the basic ideas of most quantum Hall theories \cite{Hajdu1994a}, disorder should buffer these oscillations and lead to a smooth variation of the Fermi-energy. The implications of these observations are profound, since, in the view of a closed system, jumps of the chemical potential would prevent the formation of Hall plateaus \cite{Hajdu1994a} (see Figure~\ref{fig:fluc}). Disorder must be invoked to allow smooth variation of the Fermi energy. From the viewpoint of an open system, these fluctuations are correlated with the intersections of the classical and quantum mechanical Hall resistivities. The excess charges lead also to a varying electrostatic potential above the Hall system. In an open system, disorder is not an essential ingredient of plateau formation. A careful analysis of these intersection points and their matching to the observed fluctuations is crucial to investigate this issue.

\subsection{Fermi energy and Hall potential variations}

\begin{figure}[t]
\centerline{\includegraphics[width=0.65\textwidth]{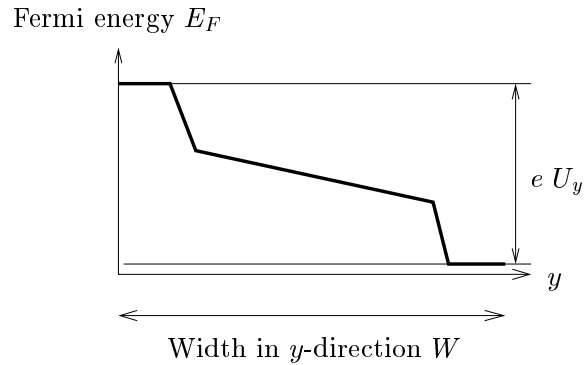}}
\caption{Schematic sketch of a possible variation of the Hall potential and the Fermi energy throughout the sample according to \cite{Ahlswede2002a}.}
\label{fig:HallPotential}
\end{figure}
Up to now we considered only the injection of electrons from a single point source. We will now extend the formalism to cover a continuous ``wire'' of point sources along the current injecting contacts. Also, we will introduce the possibility of a local variation of the Fermi energy $E_F(y) $and the Hall field $\Elf_y(y)$, as sketched in Figure~\ref{fig:HallPotential}. The current of a macroscopic device with width $W$ is given by integrating the current density over the width of the device:
\begin{equation}
I_x = \int_0^W\rmd y\;j_x(x,y) = \int_0^W \rmd y\; \sigma_{xy}\Elf_y(y) \;.
\end{equation}
For a given form of the Hall potential and the Fermi energy we are now in a position to calculate the resistivity $\boldsymbol\rho$.

\subsection{Calculation of the Hall resistivity and the current flow}

For the realistic calculation of a Hall resistivity curve we have to incorporate some material parameters in the theory:
\begin{enumerate}
\item Electrons in solids are characterized by an effective mass $m^*$.
\item Similarly the magnetic $g$-factor of the electron depends on the material and possibly on the magnetic field $\mgf$.
\item In some materials additional degeneracies appear (e.g.\ the ``valley splitting'' in silicon).
\item All observations are made at a finite temperature $T$.
\item The electric field and Fermi energy may vary along the direction of the Hall field $\Elf_y$.
\item In a multi-electron system Coulomb interactions between the electrons and the positive background charges occur.
\item In a non-perfect sample disorder and electron-phonon interactions are present.
\end{enumerate}

\subsubsection{A new expression for the  Hall conductivity}

The combination of the quantum source model with the Pauli principle allows us to obtain a purely quantum mechanical expression for the current along the drift direction. Working in the eigenfunction expansion for the Green function (see Section~\ref{subsec:DOS2D}), each eigenstate supports the current
\begin{equation}
j_x^{n,y_c}(\mbfr)
=\frac{\hbar}{m}
\Im\left\{\psi_{n,y_c}(\mbfr)^*\partial_x\psi_{n,y_c}(\mbfr)\right\}
+\frac{e A_x}{m}{|\psi_{n,y_c}(\mbfr)|}^2.
\end{equation}
The properly weighted current is given by
\begin{equation}
j_x(\mbfr;E)=
\sum_{n=0}^\infty\int\rmd y_c\;\delta [E-E_n(y_c)]
j_x^{n,y_c}(\mbfr).
\end{equation}
A short calculation yields
\begin{eqnarray}
j_x(\mbfr;E)
&=&
v_D\sum_{n=0}^\infty\int\rmd y_c\;\delta [E-E_n(y_c)]\;
{|\psi_{n,y_c}(\mbfr)|}^2\nonumber\\
&&+
\frac{e\mgf}{m}\sum_{n=0}^\infty\int\rmd y_c\;\delta [E-E_n(y_c)]\;
(y_c-y)\;{|\psi_{n,y_c}(\mbfr)|}^2.
\end{eqnarray}
We already evaluated the first term in (\ref{eq:crossed33}). A (macroscopic) conductivity is obtained by integrating over $y$. Since the second integral runs over a function antisymmetric in $(y-y_c)$, the second term vanishes. Inserting $v_D=\Elf/\mgf$ gives
\begin{equation}
j_x(\mbfr;E)=\frac{e\Elf}{\mgf} n_{\ELF\times\MGF}^{(2D)}(\mbfr;E).
\end{equation}
Without integration, the second term is responsible for the complicated flow pattern seen in the current pictures analyzed in Section~\ref{sec:qmdrift}. In a many electron system with Fermi energy $E_F$ the conductivity in Ohms law $\mbfj=\mbfsigma\cdot\mbfE$ becomes
\begin{eqnarray}
\sigma_{xy}&=&
\frac{e}{\mgf }\int_{-\infty}^{E_F}\rmd E\;
n_{\ELF\times\MGF}^{(2D)}(\mbfr;E)
=\frac{e}{\mgf }N(\mbfr;E_F),\\
\sigma_{xx}&=&0.
\end{eqnarray}
We emphasize that this expression couples the specific form of the density of states in crossed electromagnetic fields with the drift velocity. It is not possible to separate the quantity $N(\mbfr;E_F)$ from the drift velocity and introduce it as an independent classical parameter. 

\subsubsection{A simple model for the quantum Hall effect including scattering}

The previous model is not complete, because the longitudinal resistance is always zero. Experiments show a non-vanishing $\sigma_{xx}$ if $E_F$ coincidences with a Landau-level. A natural extension of the model is the incorporation of scattering. A simple, yet instructive model for the Hall effect that incorporates effects 1--4 is presented in \cite{Kramer2003b,Kramer2003a}. We start from a conductivity tensor similar to the classical expression (\ref{eq:sigmaclassical}). However, we take into account the external field and energy dependence of all quantities:
\begin{equation}
\mbfsigma(E)=\frac{e^2 n(\mbfo;E)\tau(E)}{m}
\frac{1}{1+\omega_C^2\tau(E)^2}
\left(\begin{array}{cc}
1&-\omega_C\tau(E) \\ \omega_C\tau(E)&1
\end{array}\right).
\end{equation}
We will not consider a locally varying electric field. The discussion also assumes that the electrons are injected at a point-contact located at $\mbfo$. We can lift this restriction by introducing a position dependent Fermi-level \cite{Halperin1986a} in the system
\begin{equation}\label{eq:EFtilt}
E_F(\mbfr)=E_F(\mbfo)+e\;\mbfr\cdot\Elf.
\end{equation}
Usually, unequal Fermi levels result in a current. In the Hall geometry this current along the electric field is absent as the electrons only drift perpendicular to the magnetic and electric fields. From the symmetry relation~(\ref{eq:EBGauge}) we obtain a translational invariance in the sense that
\begin{equation}
n(\mbfr;E_F(\mbfr))=n(\mbfo;E_F(\mbfo)).
\end{equation}
Therefore we can regard the local density of states as the global density of states in the system. A more sophisticated model, which incorporates a (slow) variation of the electric fields and the Fermi energy, is sketched in Section~\ref{sec:FlowPattern}. In the Appendix we show, that the Lorentz-force model for the conductivity may be replaced by the expression for the probability current defined in (\ref{eq:JDelta}).

For $T\rightarrow 0$, the total conductivity is obtained by integrating over the occupied energy range
\begin{equation}
\mbfsigma=\int_{-\infty}^{E_F}\rmd E\;\mbfsigma(E).
\end{equation}
For strong magnetic fields, the energy-dependent relaxation time $\tau(E)$ satisfies
${[\omega_C\tau(E)]}^2 \gg 1$, except in the vicinity of $E_F$, and the transversal component $\sigma_{xy}$ thus mirrors the integrated density of states $N(\mbfo;E_F)$ (\ref{eq:IDOS})
\begin{equation}
\label{eq:sigmaxy}
\sigma_{xy}=
\frac{e}{\mgf }\int_{-\infty}^{E_F}\rmd E\;
\frac{n(\mbfo;E)}{1+[\omega_C\tau(E)]^{-2}}
=\frac{e}{\mgf }N(\mbfo;E_F)
\end{equation}
[cf.~(\ref{eq:rho_cl})]. The last expression is already known from our first model: The quantization of the plateaus in the Hall effect does not depend on the scattering.

The longitudinal component is more difficult to evaluate since it involves assumptions about the scattering events. If we assume that only electrons with energies close to the Fermi energy contribute significantly to
$\sigma_{xx}$, we obtain:
\begin{equation}
\sigma_{xx}(E_F) = \frac{e}{\mgf }\int_{-\infty}^{E_F} \rmd E\;
n(\mbfo;E)\frac{\omega_C\tau(E)}{1+\omega_C^2\tau(E)^2}
\approx
D n(\mbfo;E_F).
\end{equation}
Here $D$ denotes some constant that may depend on the material parameters and the fields. For finite temperature, significant scattering may also take place in an energy range of several $k_B T$ around $E_F$. The value of $\sigma_{xx}(T)$ then follows after suitable averaging \cite{Wei1985a}
\begin{equation}
\label{eq:sxx_t}
\sigma_{xx}(T)=\int\rmd E\;\left(-\frac{\partial f(E,T)}{\partial E}\right)\;\sigma_{xx}(E),
\end{equation}
where $f(E,T)$ denotes the Fermi-Dirac distribution
\begin{equation}
f(E,T)=\frac{1}{\rme^{(E-E_F)/(k_B T)}+1}.
\end{equation}

\paragraph*{The Hall resistivity for a fixed magnetic field}

\begin{figure}[t]
\centerline{\includegraphics[width=0.8\textwidth]{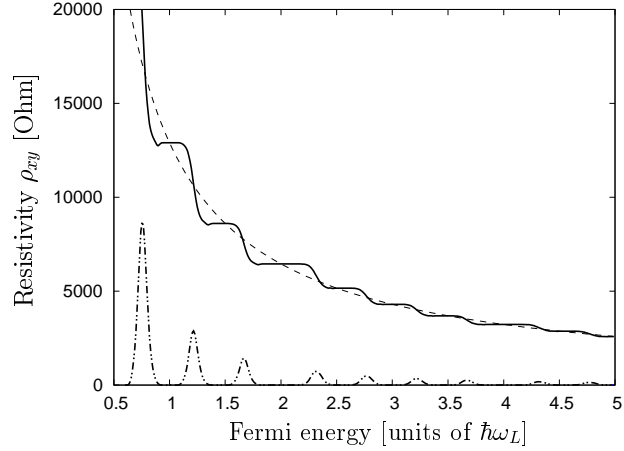}}
\caption{Quantum Hall effect for a constant magnetic field $\mgf=19$~T. Effective mass $m^*=0.2 m_e$, $T=1.5$~K. Notice that any substructure of the Landau levels is washed out by thermal averaging. The dashed line shows the classical Hall line. The lower dash-dotted line is proportional to the longitudinal resistance $\rho_{xx}$. Corresponding experimental data is shown in \cite{Klitzing1980a}.}
\label{fig:klitzing}
\end{figure}
In early experiments on the quantum Hall effect, the resistivity was measured under the condition of a fixed current along the $x$-axis ($I_x={\rm const}$) and a fixed magnetic field \cite{Klitzing1980a}.
By varying the gate voltage in the experimental Si--MOSFET system, the Fermi energy is adjusted. We will assume a linear relationship between the gate voltage and the Fermi energy. Then, the value of the Hall field $\Elf_y$ is a solution of the implicit equation
\begin{equation}
\Elf_y=\rho_{xy}(E_F,\Elf_y,\mgf)\;j_x
\end{equation}
for given $E_F, \mgf, j_x$. Here, the resistivity $\rho_{xy}$ is related to the conductivity components (\ref{eq:sigmaxy}), (\ref{eq:sxx_t}) via (\ref{eq:qh00}).  For the interpretation of data, we have to include an additional degeneracy besides spin that occurs in silicon, the ``valley splitting'' which effectively doubles each level, leading to a total of four repetitions of each Landau level. Introducing the additional valley quantum number $v=\pm\frac{1}{2}$ and the corresponding energy shift $E_v$ \cite{Klitzing1981a}, the density of states given in equation~(\ref{eq:spin}) becomes
\begin{equation}
\label{eq:spinvalley}
n_{\uparrow,\downarrow,{\rm valley}}(\mbfo;E) =n\left(\mbfo;E \pm g\hbar\omega_L/2\pm v E_v\right).
\end{equation}

\paragraph*{Temperature dependence}
If the ratio $k_B T/(\hbar\omega_L)$ becomes close to unity, the Hall plateaus disappear, since $\sigma_{xx}(T)$ as given by equation~(\ref{eq:sxx_t}) is no longer approaching zero between two Landau levels. However, the introduction of an effective mass $m^*$ can lead to large modifications of $\omega_L^*=eB/(2m^*)$. Since the width $k_B T$ is independent of material parameters, it can be used as an independent energy scale to access the values of $m^*,g^*$. In Figure~\ref{fig:klitzing} we assume an effective mass  $m^*/m_e=0.2$. A higher effective mass would be inconsistent with the reported temperature of $T=1.5$~K, since a smaller energy range $\hbar\omega_L^*$ cannot accommodate four separated peaks of individual width $k_B T$. Thus, the temperature dependence of $\rho_{xy}$ may be used to determine some of the material parameters of the system.

\begin{figure}[t]
\centerline{\includegraphics[width=0.9\textwidth]{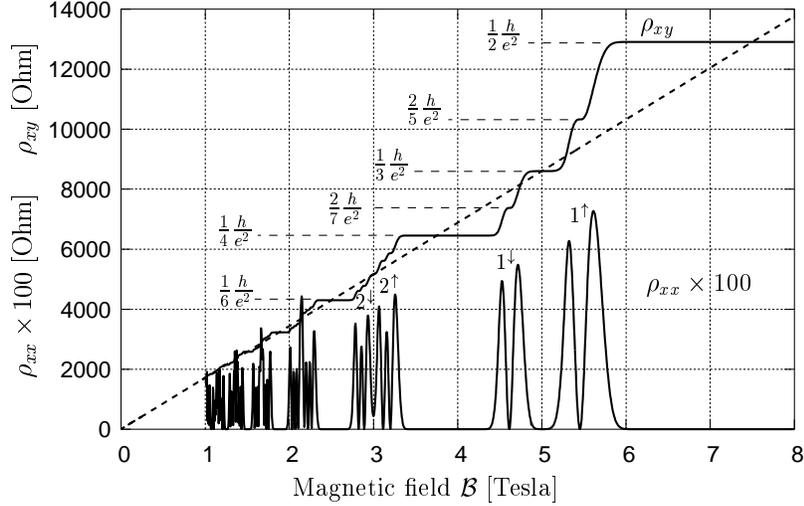}}
\caption{Quantum Hall effect at strong magnetic fields ($B>1$~Tesla) for a non-interacting two-dimensional electron gas.  The plot shows the Hall resistance $\rho_{xy}$ and longitudinal resistance $\rho_{xx}$ as a function of the magnetic field $\mgf$ for fixed Fermi energy ($E_F=0.868$~meV).  Effective mass $m^*=m_e$, effective $g$-factor $g=\frac12$, current density $j_x=0.2$~A/m, $\tau(E_F)=10^{-11}$~s, $T=0.1$~K. The dashed line represents the classical Hall resistance $\rho_{xy}$ with a constant level density. Experimental results are shown in \cite{Paalanen1982a}. Note that $j_x$ is chosen fairly large in order to show the substructure of the Landau levels.}
\label{fig:HallB}
\end{figure}

\paragraph*{The Hall resistivity as a function of the magnetic field}

Nowadays, GaAs--heterostructures are commonly used to provide the two-di\-men\-sional electron gas for the quantum Hall effect. Advantages are cleaner samples with very high mobilities and the absence of valley splittings. However, in these samples the number of electrons for $\mgf=0$ is virtually constant and largely independent of the gate voltage. Therefore the magnetic field is varied while keeping the gate voltage fixed. In Figure~\ref{fig:HallB} we show a typical plot of the resulting resistivity. As mentioned before, the intersection points of the averaged or classical Hall resistivity with the quantum mechanical result for $\rho_{xy}$ deliver important information: They allow to analyze the spin splitting and other parameters of the system.

\subsubsection{Fractional effects}

As shown before, the electric field leads to additional zeroes in the density of states and consequently to subdivided Landau levels. Their fractional values of the filling factor are analyzed in \cite{Kramer2003a} and displayed in Figure~\ref{fig:HallB}. Experimentally, the appearance of the fractional filling factors arising from the electric field might be difficult to detect, since uniform and high current densities are required at very low temperatures. Furthermore, the fractional quantum Hall effect may overshadow the single-particle structure if Coulomb interactions dominate the Hall field contribution. Experiments show plateaus in $\rho_{xy}$ for simple fractions of the filling factor in the first ($n=0$) Landau level. In the presented model such features are not explained. The effect is attributed to collective modes of the system that are caused by interactions between the electrons \cite{Yoshioka2002a}. We explicitly did not include Coulomb interactions in our model system.

The best candidates for the detection of field-induced fractional filling factors are exactly half-filled odd Landau levels, since they are left largely unaffected by the averaging caused by a non-uniform electric field. (A similar effect prevails in three dimensions, where the dip associated with the second Landau level is still visible in Figure~\ref{fig:Baverage} despite extensive averaging over different field strengths.)  In contrast, according to the standard theory of the fractional quantum Hall effect half-filled levels do not induce plateau formation.

\subsubsection{Hall-field dependence of the plateau width}

\begin{figure}[t]
\centerline{\includegraphics[width=1.0\textwidth]{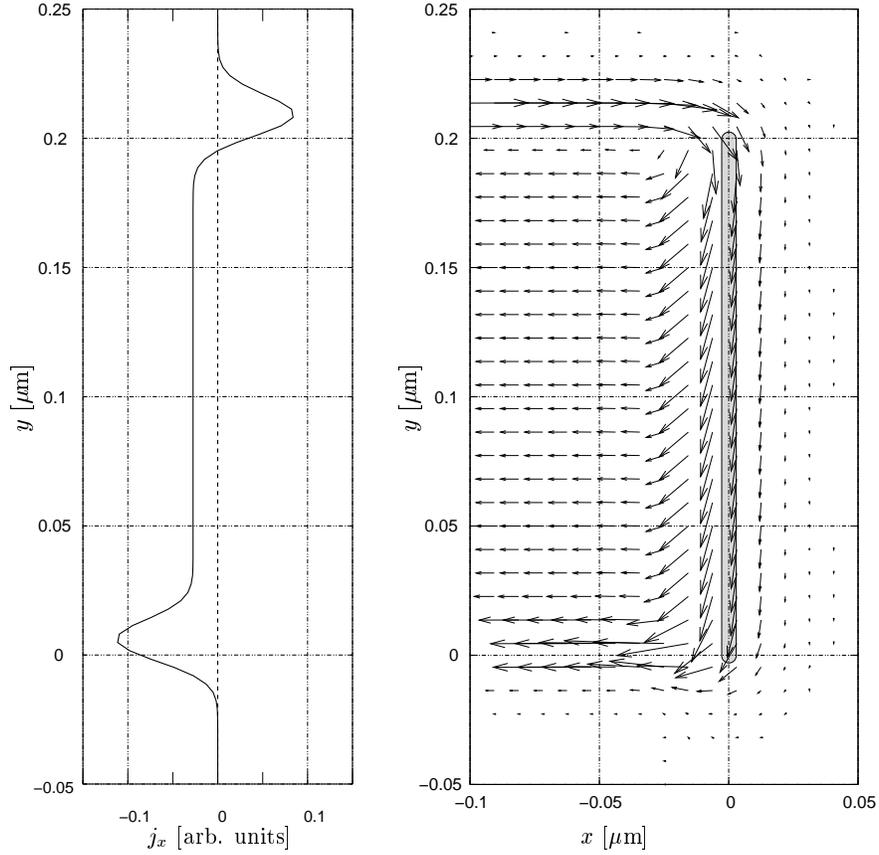}}
\caption{Quantum mechanical current density from an extended ``wire'' (shaded region). The electron emitting region extends from $y=0$ to $0.2\;\mu$m. Left panel: The component $j_x(x,y)$ at $x=-0.1\;\mu$m. Right panel: Spatial current flow. The arrows indicate the direction of the current and their length is proportional to $|j|^{1/4}$. In this example, an effective edge current forms with unequal magnitudes at each edge. The bulk also carries a constant current flow.}
\label{fig:jzextended}
\end{figure}
Another signature of the presence of an electric field dependent broadening of Landau levels is the  breakdown behavior for high currents. Kawaji and co-workers conducted extensive experiments on the characteristics of the breakdown and find experimentally a dependency that is exactly the same as the one obtained in equation~(\ref{eq:LLoverlap}): The width of the plateaus decreases linearly with increasing current. The experimental observation of this behaviour is reported in \cite{Bliek1986a,Kawaji1993a,Shimada1998a}.

\subsection{Current distribution}\label{sec:FlowPattern}

The form of the Hall potential is actually experimentally accessible \cite{Ahlswede2002a}, and a schematic result is sketched in Figure~\ref{fig:HallPotential}. We should note that the theory presented here could be applied for any given variation of the Hall potential and the Fermi energy. A starting point for models that provide this input could be the self-consistent potentials obtained in Refs.\ \cite{Chklovskii1992a,Lier1994a}. For the sake of simplicity, let us discuss here only a straightforward extension in which we treat the emission of independent electrons along a constant Hall field.  We already calculated the current distribution for a point source. For the ``wire'' described above, we obtain the global current profile by summing over the current contributions of the point sources. Here, we will use eq.~(\ref{eq:EFtilt}) and assume that the Hall field is constant across the probe.

Figure~\ref{fig:jzextended} displays the resulting flow pattern. The previously described oppositely flowing currents are shifted to the edges, while in the bulk a uniform current emerges. In this way effective ``edge'' currents are established in a model of an electron emitting contact of finite width. We should note that the magnitudes of the oppositely directed edge currents differ, as already seen for point sources.

\section{Conclusions}

The quantum source formalism provides an excellent basis for the analysis of the propagation of matter waves in external fields. While some classical properties of the motion of particles prevail in the quantum mechanical case, a smooth transition from quantum to classical mechanics is generally not observed. The Landau quantization due to the magnetic field and the combination with an electric field have profound implications for the spatial current distribution and intensity.

Present-day nanotechnological devices, operated at very low temperatures, can actually provide experimental data for the propagation of electronic matter waves and allow to test theoretical predictions. We analyzed a simple model of the quantum Hall effect as one example. Even this non-interacting electron picture already shows a wealth of interesting features and gives access to important parameters of the system.

\begin{acknowledgements}
T.~K.\ would like to thank Prof.\ Bruno Gruber for the invitation to present this paper at the ``International symposium symmetries in science XIII''. This work was financially supported by the Deutsche Forschungsgemeinschaft (project number Kl315/6-1, T.~K.), and the Alexander von Humboldt foundation and the Killam trust (C.~B.).
\end{acknowledgements}

\end{article}
\end{document}